\documentclass[twocolumn]{aastex62}

\usepackage{amsmath}
\usepackage{amssymb}
\usepackage{graphicx}
\usepackage{epstopdf}
\usepackage{inputenc}
\usepackage[toc]{appendix}
\usepackage{xcolor}
\usepackage{hyperref}

\usepackage{amsmath}

\def \sun {$_{\odot}$}

\graphicspath{{./}{}}

\received{}
\revised{\today}
\accepted{}
\submitjournal{ApJ}

\shorttitle{[C\,II] kinematics of IC\,59 and IC\,63}
\shortauthors{M. Caputo et al.}

\begin{document}

\title{Physics and Chemistry of Radiation Driven Cloud Evolution. [C\,II] Kinematics of IC\,59 and IC\,63}

\correspondingauthor{Miranda Caputo}
\email{miranda.caputo@rockets.utoledo.edu}

\author[0000-0002-2957-3924]{Miranda Caputo}
\affil{Ritter Astrophysical Research Center, University of Toledo Dept. of Physics and Astronomy, 2801  W. Bancroft St. Toledo, Ohio 43606, USA}
\affil{SOFIA Science Center, USRA, NASA Ames Research Center, M.S.-12, N232, Moffett Field, CA 94035, USA}
\author[0000-0002-6386-2906]{Archana Soam}
\affil{Indian Institute of Astrophysics, II Block, Koramangala, Bengaluru 560034, India}
\affil{SOFIA Science Center, USRA, NASA Ames Research Center, M.S.-12, N232, Moffett Field, CA 94035, USA}
\author[0000-0001-6717-0686]{B-G Andersson}
\affil{SOFIA Science Center, USRA, NASA Ames Research Center, M.S.-12, N232, Moffett Field, CA 94035, USA}
\author{Remy Dennis}
\affil{Santa Clara University, 500 El Camino Real, Santa Clara, CA 95053, USA}
\author[0000-0003-4195-1032]{Ed Chambers}
\affil{SOFIA Science Center, USRA, NASA Ames Research Center, M.S.-12, N232, Moffett Field, CA 94035, USA}
\author[0000-0002-1708-9289]{Rolf G\"usten}
\affil{Max Planck Institut f\"ur Radio Astronomie, Bonn, Germany}
\author[0000-0002-9342-9003]{Lewis B.G. Knee}
\affil{Herzberg Astronomy and Astrophysics Research Centre, National Research Council of Canada, 5071 West Saanich Road, Victoria, BC, V9E 2E7, Canada}
\author[0000-0001-7658-4397]{J\"urgen Stutzki}
\affil{I. Physikalisches Institut der Universit\"{a}t zu K\"{o}ln, Z\"{u}lpicher Straße 77, D50937 K\"{o}ln, Germany}



\begin{abstract}

We used high-resolution [C\,II] 158\,$\micron$ mapping of two nebulae IC\,59 and IC\,63 from SOFIA/upGREAT in conjunction with ancillary data on the gas, dust, and polarization to probe the kinematics, structure, and magnetic properties of their photo-dissociation regions (PDRs). The nebulae are part of the Sh 2-185 H\,II region illuminated by the B0 IVe star $\gamma$ Cas. The velocity structure of each PDR changes with distance from $\gamma$ Cas, consistent with driving by the radiation. Based on previous FUV flux measurements of, and the known distance to $\gamma$ Cas along with the predictions of 3D distances to the clouds, we estimated the FUV radiation field strength (G$_0$) at the clouds. Assuming negligible extinction between the star and clouds, we find their 3D distances from $\gamma$ Cas. For IC\,63, our results are consistent with earlier estimates of distance from  \citet{2013ApJ...775...84A}, locating the cloud at $\sim$2\,pc from $\gamma$ Cas, at an angle of 58$^\circ$ to the plane of the sky, behind the star. For IC\,59, we derive a distance of 4.5\,pc at an angle of 70$^\circ$ in front of the star. We do not detect any significant correlation between the orientation of the magnetic field \citep{2017MNRAS.465..559S} and the velocity gradients of [C\,II] gas, indicating a moderate magnetic field strength. The kinetic energy in IC\,63 is estimated to be order of ten higher than the magnetic energies. This suggests that kinetic pressure in this nebula is dominant.

\end{abstract}

\keywords{ISM: Clouds – Submillimeter: ISM – ISM: lines and bands – ISM: individual (IC\,63) – ISM: individual (IC\,59) – (ISM:) photon-dominated region (PDR)}

\section{Introduction} \label{sec:intro}

\begin{figure}
\centering
\resizebox{8.5cm}{8cm}{\includegraphics{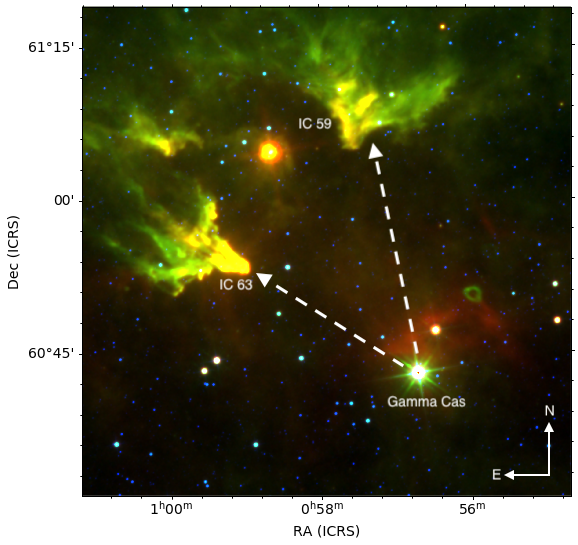}}
\caption{RGB image of the Sh 2-185 H\,II region made from WISE W1 (3.4$\micron$) in blue, W3 (12$\micron$) in green, and W4 (22$\micron$) in red.}\label{fig:ccsh2}
\end{figure}

Radiation from newly formed, hot, high-mass O and B stars can give rise to a number of dynamical and chemical effects in surrounding material. Ionizing radiation from these young stars can cause exposed material in the surrounding cloud to photoevaporate and ionize the inner boundaries of expanding HII regions. Existing density enhancements on the peripheries of expanding H\,II regions create  bright-rimmed clouds (BRCs). The bright rim produced by the recombination radiation from the ionization front (IF) on their surface facing the ionizing source.

High energy radiation from ionising sources causes feedback processes in the surrounding medium. This feedback can be positive or negative; causing triggered star formation or quenching of star formation respectively. A systematic investigation is needed on injection of mechanical energy and radiative heating efficiency in regions dominated by the different processes (stellar winds, thermal expansion, radiation pressure). 
However, some numerical simulations, such as \citet{2021ApJ...922L...3L}, shed controversy on the importance of stellar winds, leaving mainly
thermal expansion and photoevaporation. The [C\,II] line uniquely probes the kinematics of the gas exposed to the mechanical energy input by massive stars and therefore directly measures the mechanical energy injection into the medium. The observations of the [C\,II] line suggest that that winds could dominate for early type stars while thermal
expansion dominates for later type stars, e.g. \citet{2020A&A...639A...2P}. 
The dominant processes in PDRs depends not only on the stellar spectral type, but also the evolutionary state of the system, internal pressures, and the geometry and distance of the clouds from the star. 
The case of IC\,59 and IC\,63 presents an interesting example in which the feedback from a B0 star is dominated by photevaporation. 
The radiative feedback processes acts directly on the PDR while mechanical feedback is transmitted through the PDR to the cloud interiors.

Associated with these H\,II regions and BRCs, photo-dissociation regions, or photon-dominated regions (PDRs), can be described as the regions in the interstellar medium (ISM) where UV radiation dominates the photochemistry \citep{1997ARA&A..35..179H, 2022ARA&A..60..247W}. 
These PDRs are found in the transitional regions between the ionized, atomic material and the colder, denser, molecular material in clouds. UV radiation can infiltrate the denser gas in these PDR regions exciting, ionizing, and affecting the gas motion, inducing chemical, thermal, and dynamical changes, and, possibly, triggering star formation.  These features make PDRs important laboratories for probing the physics, chemistry, and evolution of the ISM.

Amongst the nearest of well-defined systems of PDRs is the Sh 2-185 H\,II region created by the B0 IVe star $\gamma$ Cas (Figure \ref{fig:ccsh2}), located at a distance of $190\pm20$pc, consisting of the IC\,63 and IC\,59 nebulae \citep{1957ApJ...125..622O,2005ApJ...628..750F,2010ApJ...717..658M,2013ApJ...775...84A,2017MNRAS.465..559S}. Because of its brightness and proximity to the sun, Sh 2-185 is an ideal laboratory to better test the modern theoretical PDR models and to understand their structure and kinematics.

The brighter of these clouds, IC\,63, has been observed through various studies and surveys and is composed of several clumps with varying characteristics. 
\citet{1994A&A...282..605J,1995A&A...302..223J,1996A&A...309..899J,1996A&A...315..327J} performed an extensive, multi-line tracer survey of IC\,63 and interpreted their relatively low spatial resolution data with comprehensive physical and chemical modeling without resolving the structure of the cloud itself. 
Using interferometric maps in $\rm HCO^{+}$\,(J=1-0), \citet{2005IAUS..231P.148P} showed a highly compressed ridge on the star facing side of IC\,63 which is consistent with numerical models of BRCs \citep{2009ApJ...692..382M}. \citet{2010ApJ...725..159F} used $\rm H_{2}$\,(0-0)\,S(2) through S(5) excitation to show a temperature gradient across the cloud. Due to the shape of IC\,59 not being represented by most BRC models \citep{2009ApJ...692..382M}, \citet{2010ApJ...717..658M} proposed and modeled an idealized  M-type BRC morphology based on IC\,59 structure and supported the conclusion from \citet{2005AJ....129..954K} that there is no active star formation in IC\,59. 

\citet{2018A&A...619A.170A} used infrared emission from both IC\,63 and IC\,59 to study the PAHs in these PDRs. They concluded that both IC\,63 and IC\,59 are experiencing photo-evaporation due to radiation from $\gamma$ Cas and that the tips of each nebula are further away from $\gamma$ Cas than their respective projected distances. They also used archival $\textit{Herschel}$ and $\textit{Spitzer}$ maps and line spectroscopy of IC\,63 and IC\,59, using spectral energy distributions to measure the FUV radiation field strength. They performed PDR modeling of the [C\,II] and [O\,I] lines and the infrared continuum to derive the density and temperatures of these clouds.

Along with surveys of gas and dust tracers, another of the many ways in which these clouds have been studied is through starlight polarization observations to investigate magnetic fields. PDRs are an ideal environment for the study of grain alignment mechanisms, in particular the radiative alignment torque (RAT) theory \citep{2007MNRAS.378..910L,2013ApJ...775...84A,2015MNRAS.448.1178H}. 
\citet{2017MNRAS.465..559S} mapped the magnetic fields in a region containing IC\,63 and IC\,59 using optical polarization measurements. The magnetic fields were found to be ordered. In IC\,63, the field lines were found almost parallel to the direction of illuminating radiation and almost perpendicular in case of IC\,59. These investigations were done to understand the magnetized evolution of these nebulae. Polarization efficiency and collisional disalignment of dust grains in IC\,63 are studied by \citet{soam_poleff2021} and \citet{soam_coll2021}, respectively. The gradients in temperature and densities in the PDR region of IC\,63 are also studied by \citet{soam_exes2021} using pure rotational molecular hydrogen observations.

High sensitivity observations of a PDR line-tracer at high spectral resolution is necessary in order to probe the kinematics of the region. SOFIA upGREAT velocity maps of [C\,II] are ideal for tracing PDRs. In this paper, our focus is to understand the kinematics, structure, and geometry of these nebulae using [C\,II] emission.

The paper is structured as follows: 
Section \ref{sec:obs} describes the observations and data reduction. 
Section \ref{sec:analysis} presents the detailed analysis. The results of the gas and dust emissions and cloud kinematics, magnetic field and velocity gradients, and the FUV radiation strength estimation in IC\,63 are given in Section \ref{sec:results}. In Section \ref{sec:discussion}, we discuss our findings and Section \ref{sec:summary} summarizes our results and describes proposed further work.

\section{Data acquisition and reduction} \label{sec:obs}

We observed [C\,II] 158\,$\micron$ line emission toward IC\,63 and IC\,59 using the upgraded German Receiver for Astronomy at Terahertz Frequencies (upGREAT) \citep{2012A&A...542L...1H, 2016A&A...595A..34R} on board the Stratospheric Observatory for Infrared Astronomy (SOFIA) \citep{2012ApJ...749L..17Y}. The upGREAT Low Frequency Array (LFA) is a 14-pixel array (two 7-pixel arrays with orthogonal polarizations) that observes at $\sim$ 2\,THz. Observations of both sources were made as part of SOFIA project 05\_0052 (PI: B-G Andersson).

The IC\,63 observations were made on a $\sim$50-minute flight leg on 1 February 2017. To create the map, we used the total power on-the-fly mode (TP OTF), scanning in the N-S direction. The map size was $\rm 288^{\prime \prime}$ x $\rm 252^{\prime \prime}$, as seen by the central LFA pixel. The step size between spectral readouts along each scan row was $\rm 6^{\prime \prime}$, and the spacing between scan rows was $\rm 6^{\prime \prime}$. The integration time in each readout was 0.7 s\,pixel$\rm ^{-1}$, which translates to $\sim$10 seconds of on-source integration time, given the spatial multiplexing advantage of 14 pixels.

The IC\,59 observations were carried out during six flight legs in 2017, on 8, 9, 10, 14 February and 13 and 15 June. The total flight leg time for these six legs was $\sim$3.5 hours. The IC\,59 region was covered with two different TP OTF maps. The first, larger map is $\rm 600^{\prime \prime}$ x $\rm 540^{\prime \prime}$ as seen by the central LFA pixel, and covers the entire nebula. The second, smaller map is $\rm 468^{\prime \prime}$x$\rm294^{\prime \prime}$, and is rotated by 25$^{\circ}$ (counter-clockwise) relative to the J2000 coordinate frame. The sky position of the second map was optimized to cover the [C\,II] emission detected in the first map. Both IC\,59 maps have $\rm 6^{\prime \prime}$ between readouts along each scan row, and $\rm 6^{\prime \prime}$ separation between rows. The total on-source integration time per map point in the combined map is $\sim$14 seconds.

All of the data was calibrated by the GREAT instrument team using their \textit{kalibrate} software \citep{2012A&A...542L...4G}. Main beam efficiencies for each pixel were applied (values ranged from 0.59 to 0.68). We fit second order baselines to a narrow spectral region (-21 to +21 km\,s$\rm ^{-1}$ for both IC\,63 and IC\,59) after masking out the line region (-4 to +4 km\,s$\rm ^{-1}$ for IC\,63, -5 to +5 km\,s$\rm ^{-1}$ for IC\,59). The final data cubes were gridded onto $\rm 7.5^{\prime \prime}$ pixels, and spectrally smoothed to 0.4\,km\,s$\rm ^{-1}$. To remove spectral baselines and create the final data cube, we used  the CLASS\footnote{CLASS is part of the Grenoble Image and Line Data Analysis Software \citep[GILDAS;][]{pety2005}, which is provided and actively developed by IRAM, and is available at http://www.iram.fr/IRAMFR/GILDAS} software package.

We also used $\rm HCO^{+}$\,(J=1-0) position-position-velocity maps observed with CARMA\footnote{The Combined Array for Research in Millimeter-wave Astronomy. \url{https://en.wikipedia.org/wiki/Combined_Array_for_Research_in_Millimeter-wave_Astronomy}} \citep{soam2021b}, and molecular hydrogen $\rm H_{2}$\,(1-0)\,S(1) mapping from CFHT\footnote{Canada France Hawaii Telescope. \url{https://www.cfht.hawaii.edu/}} \citep{2013ApJ...775...84A} in conjunction with low-resolution $\rm ^{12}CO$\,(1-0) data from TRAO\footnote{Teaduk Radio Astronomical Observatory at the Korea Astronomy and Space Science Institute. \url{https://trao.kasi.re.kr/main.php}} as part of a molecular line survey of 16 BRCs by Soam et al. (in prep.) to help in studying the gas locations at different opacities. 
Visual extinction data was obtained from a Vilnius-system \footnote{VILNIUS is a medium-band, seven-color photometeric system(UPXYZVS).} Archival 70 and 250 -$\micron$ maps from \textit{Herschel} PACS and SPIRE, WISE 22\,$\micron$ maps, \textit{Spitzer} IRAC 8\,$\micron$, and IPHAS\footnote{The International Photometric H$\alpha$ Survey of the northern galactic plane. \url{http://www.iphas.org/}} H$\alpha$ \citep{2005MNRAS.362..753D,2014MNRAS.444.3230B} maps of the regions were used to better understand the dust properties of the clouds.

\section{Analysis} 
\label{sec:analysis}

\subsection{Dust and Gas Emissions}

\begin{figure}
\centering
\resizebox{8cm}{7.5cm}{\includegraphics{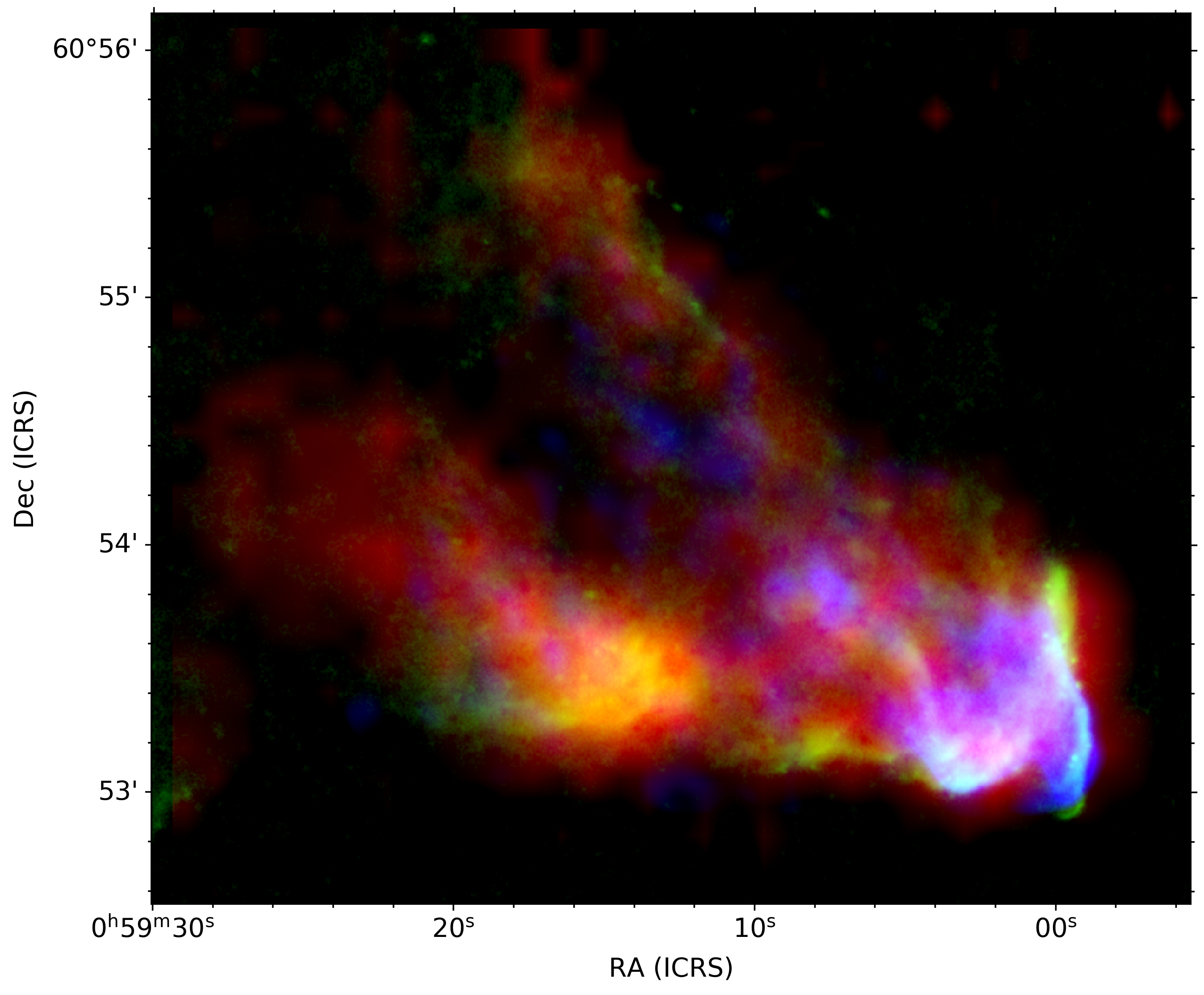}}
\resizebox{8cm}{7.5cm}{\includegraphics{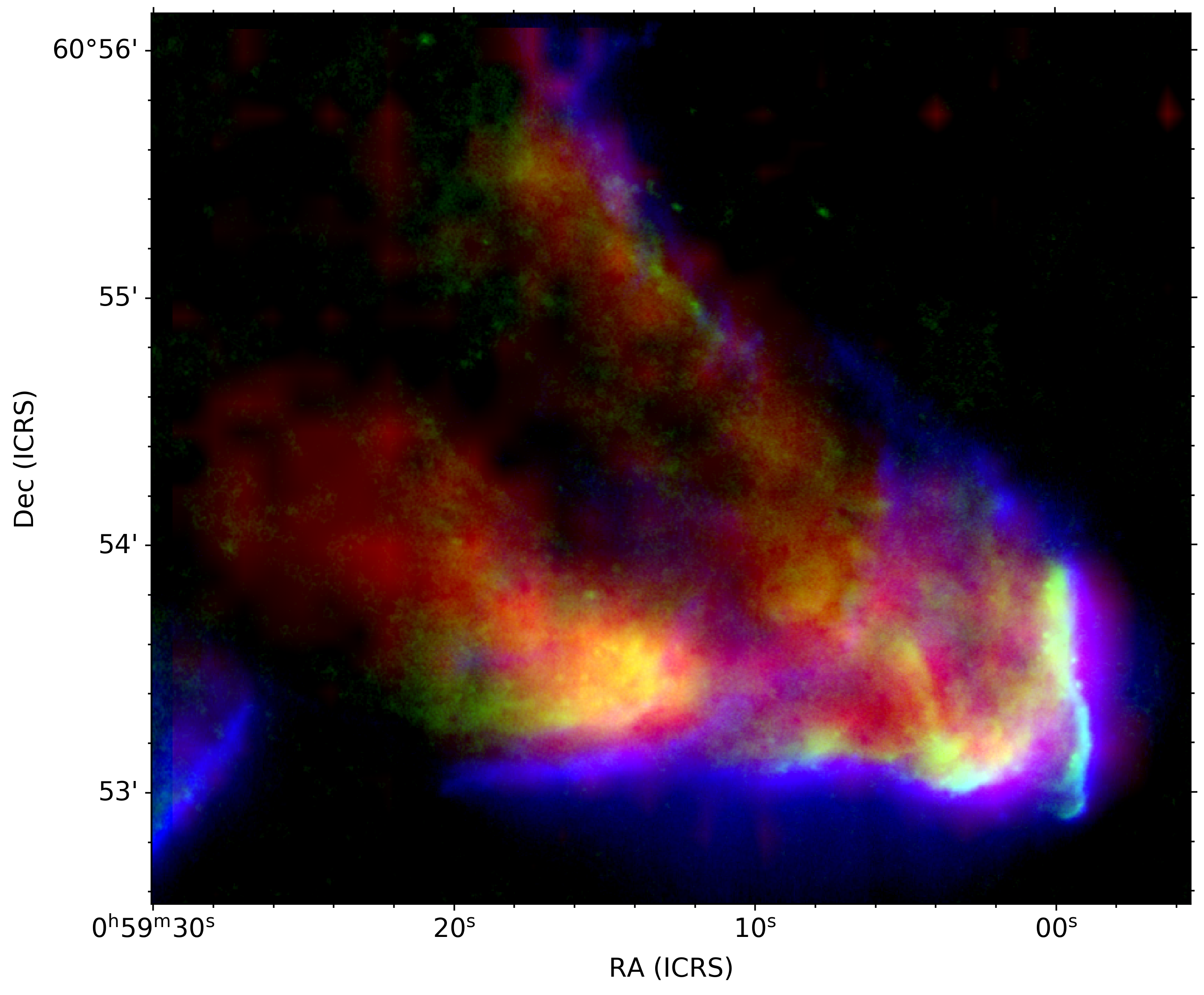}}
\resizebox{6cm}{5.5cm}{\includegraphics{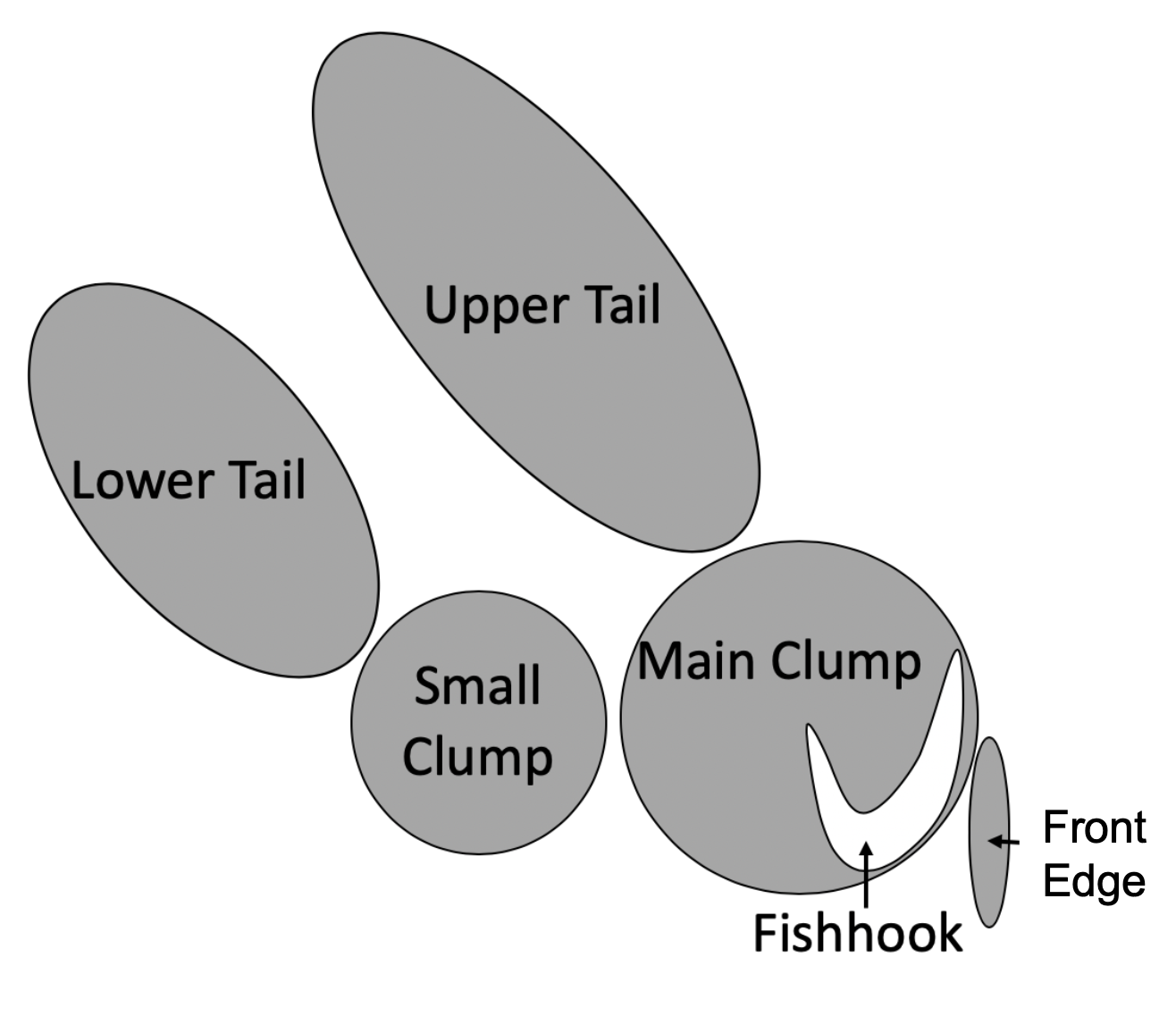}}
\caption{{\bf Top panel:} three-color, RGB image of IC\,63 made with SOFIA upGREAT [C\,II] integrated intensity map in red, CFHT $\rm H_{2}$\,(1-0)\,S(1) in green, and CARMA $\rm HCO^{+}$\,(J=1-0) integrated intensity in blue. {\bf Middle panel:}  RGB with [C\,II] in red and $\rm H_{2}$\,(1-0)\,S(1) in green like above, but blue is IPHAS H$\alpha$ emission. {\bf Lower panel:} Model of IC\,63 with naming conventions.}\label{fig:cc}
\end{figure}

\begin{figure}
\centering
\resizebox{8cm}{7.5cm}{\includegraphics{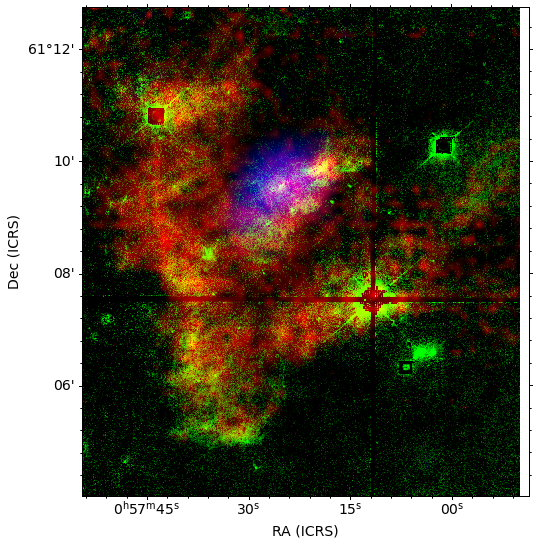}}
\resizebox{8cm}{7.5cm}{\includegraphics{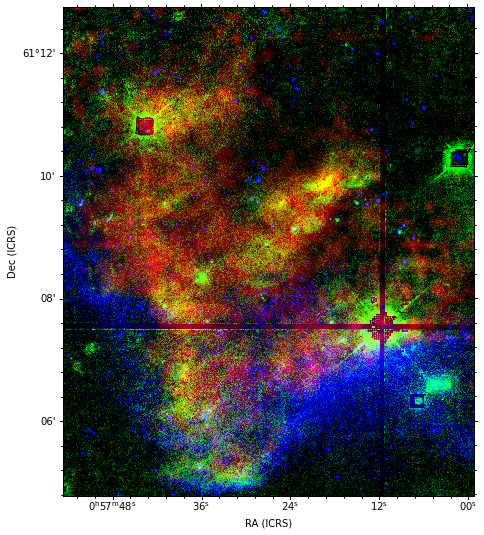}}
\resizebox{6cm}{5.5cm}{\includegraphics{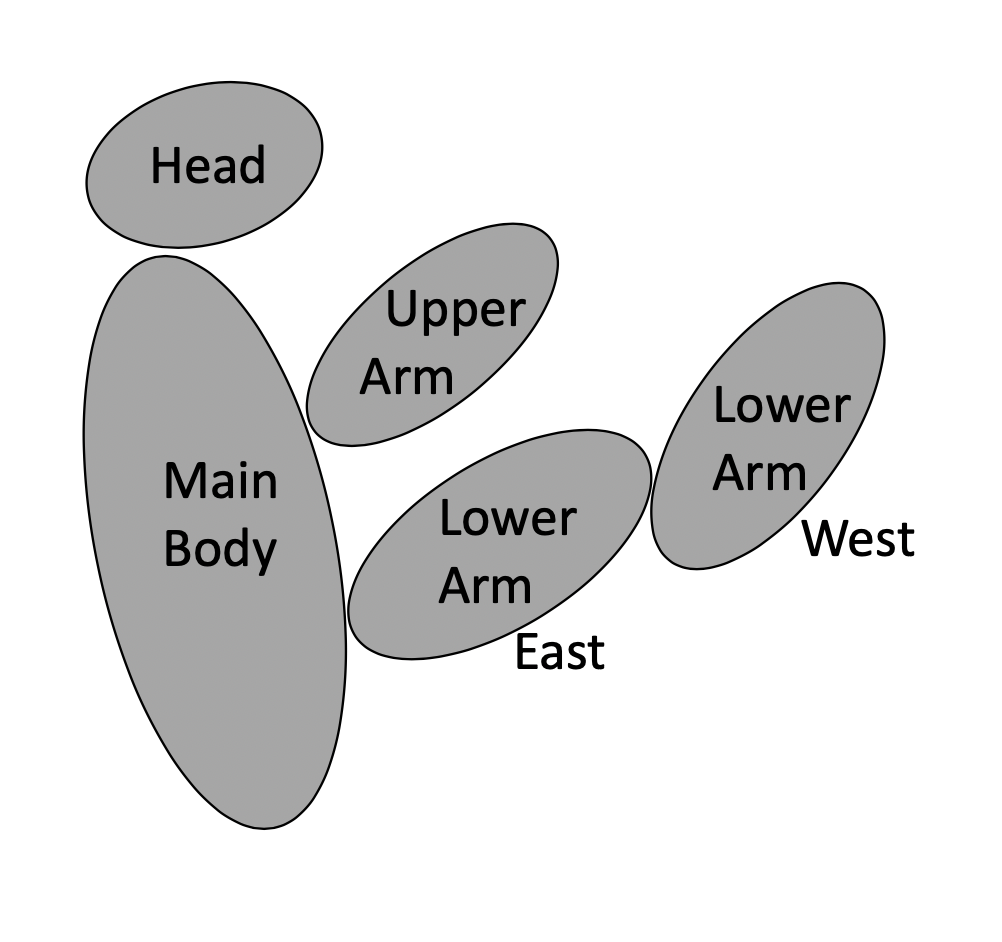}}
\caption{Same as Figure \ref{fig:cc} but for IC\,59 and the blue in the upper panel is $\rm ^{12}CO$\,(1-0) integrated intensity. Remnants of the imperfect star subtraction and chip boundaries can be seen in the CFHT $\rm H_{2}$\,(1-0)\,S(1), green, emission. }\label{fig:cc59}
\end{figure}

Figure \ref{fig:cc} shows two RGB images of IC\,63 created with the upGREAT [C\,II] integrated intensity map in red, CFHT $\rm H_{2}$\,(1-0)\,S(1) map in green, CARMA $\rm HCO^{+}$\,(J=1-0) integrated intensity map in blue in the top panel, and IPHAS H$\alpha$ emission in blue in the middle panel. 

Figure \ref{fig:cc59} shows RGB images of IC\,59 with [C\,II] in red, $\rm H_{2}$\,(1-0)\,S(1) in green, TRAO $\rm ^{12}CO$\,(1-0) integrated intensity map in blue in the upper panel and as IPHAS H$\alpha$ in blue in the middle panel.
The intensity levels for each color in Figures \ref{fig:cc} and \ref{fig:cc59} were chosen to provide a balance between signal-to-noise and the desire to best represent all the data in the images. 
The CFHT $\rm H_{2}$\,(1-0)\,S(1) and the IPHAS H$\alpha$ images were star subtracted.\footnote{Images were star subtracted thanks to the program Starnet++, found at \url{https://sourceforge.net/projects/starnet/}}. 

The lower panels in Figures \ref{fig:cc} and \ref{fig:cc59} illustrate the regions and naming conventions used in the rest of this paper for IC\,63 and IC\,59, respectively.

\begin{figure*}
\centering
\resizebox{18cm}{12cm}{\includegraphics{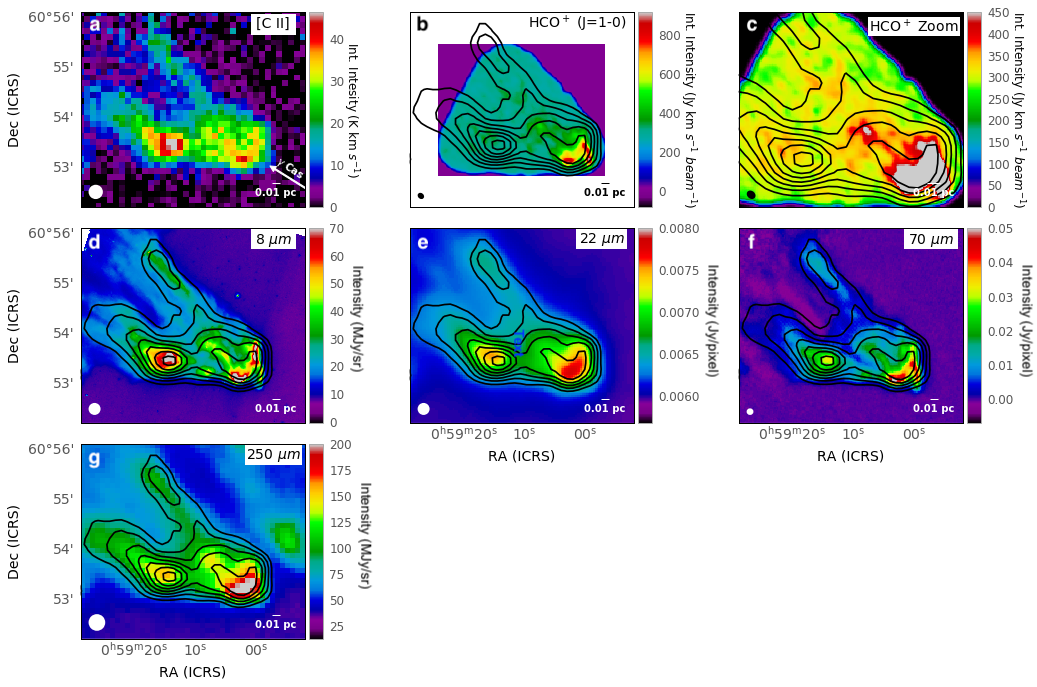}}
\caption{False color images of dust tracers and integrated intensity maps of $\rm HCO^{+}$\,(J=1-0) and [C\,II].
Dust tracer maps are \textit{Spitzer} IRAC 8\,$\micron$, WISE 22\,$\micron$, and \textit{Herschel} PACS 70\,$\micron$ and SPIRE 250\,$\micron$ maps.
Overlaid black contours show smoothed [C\,II] integrated intensity, panel (a), at levels [10, 15, 20, 25, 30, 35, 40]\,K\,km\,$\rm s^{-1}$.
Panel (c) is a zoomed in and stretched image of panel (b). 
All panels are labeled on the top right corner with the related map name, lower left with beam size, and lower right with a 0.01\,pc scale reference except for panel (c), where this is located in the upper left. 
Panel (a) also has a white arrow showing the plane of the sky projection from $\gamma$ Cas as reference.}\label{fig:dust}
\end{figure*}

\begin{figure*}
\centering
\resizebox{18cm}{13cm}{\includegraphics{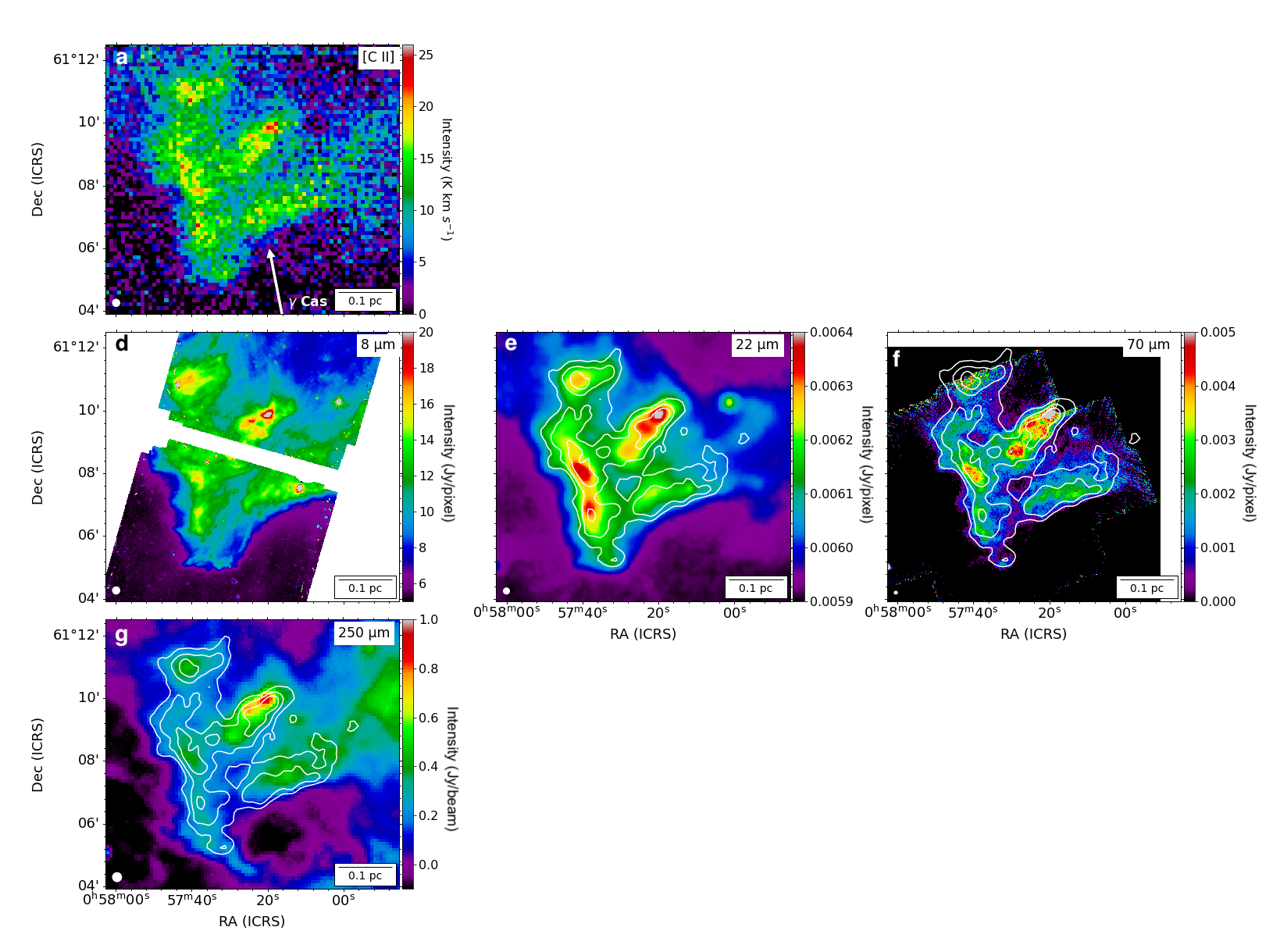}}
\caption{Same emission maps from Figure \ref{fig:dust} but for IC\,59 and without $\rm HCO^{+}$\,(J=1-0). Overlaid black and white contours show smoothed [C\,II] integrated intensity, panel (a), at levels [10, 13, 16, 20]\,K\,km\,$\rm s^{-1}$.
All panels are labeled on the lower right with a 0.1\,pc scale reference.
}\label{fig:dust59}
\end{figure*}

Figure \ref{fig:dust} shows integrated intensity maps of [C\,II] and $\rm HCO^{+}$\,(J=1-0), and dust emission maps from \textit{Herschel} PACS 70 and SPIRE 250\,$\micron$, WISE 22\,$\micron$, and \textit{Spitzer} IRAC 8\,$\micron$ for IC\,63. Panel (c) is the $\rm HCO^{+}$\,(J=1-0) integrated intensity map from panel (b) zoomed-in and with the intensities stretched in order to show the fainter emission near the second clump that is only visible when saturating the high emission region to the west. Figure \ref{fig:dust59} shows the same maps as Figure \ref{fig:dust} for IC\,59 but without $\rm HCO^{+}$\,(J=1-0) maps.

Each map from Figures \ref{fig:dust} and \ref{fig:dust59} has the corresponding beam size plotted in the lower left corner and a physical scale for reference (0.01\,pc for IC\,63 and 0.1\,pc for IC\,59) on the lower right, assuming the clouds are 200\,pc away. 
All maps other than the [C\,II] integrated intensity map are over-plotted with smoothed black or white contours of the corresponding [C\,II] integrated intensity map with levels at [10, 15, 20, 25, 30, 35, 40]\,K\,km\,s$\rm^{-1}$, $\sim$20-80\% of the maximum integrated intensity of 50\,K\,km\,s$\rm^{-1}$ for IC\,63 and [10, 13, 16, 20]\,K\,km\,s$\rm^{-1}$, $\sim$33-66\% of the maximum integrated intensity of 30\,K\,km\,s$\rm^{-1}$ in IC\,59.

\subsection{Velocity Channel Maps}

\begin{figure*}
\centering
\resizebox{15cm}{12cm}{\includegraphics{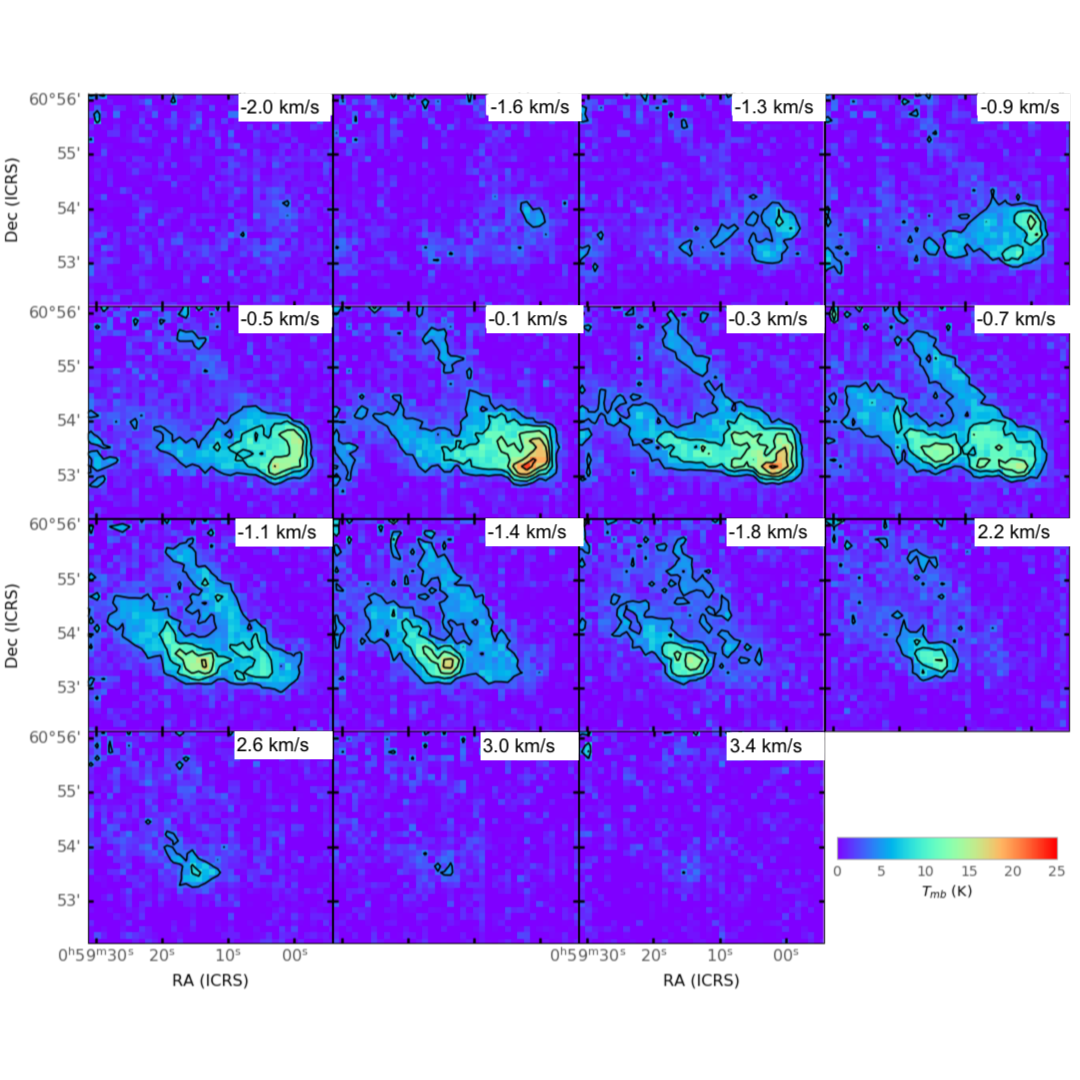}}
\caption{False color channel map of [C\,II] for IC\,63.
The corresponding velocity for all channels are marked in the upper right-hand corner of each panel.}\label{fig:cmap}
\end{figure*}

\begin{figure*}
\centering
\resizebox{15cm}{12cm}{\includegraphics{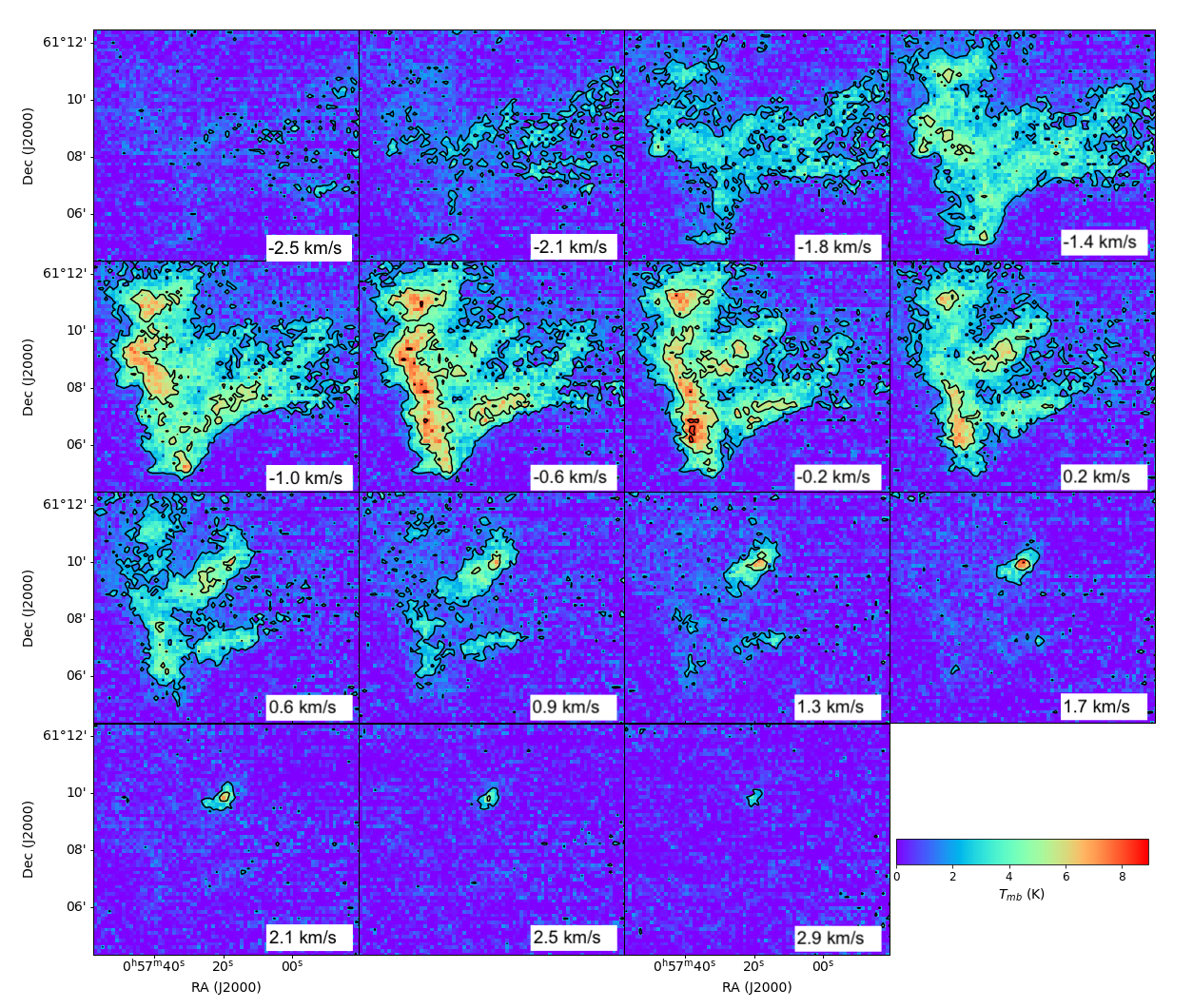}}
\caption{Same as Figure \ref{fig:cmap} but for IC\,59}\label{fig:cmap59}
\end{figure*}

Figure \ref{fig:cmap} shows a velocity channel map of [C\,II] in IC\,63 obtained from our SOFIA/upGREAT position-position-velocity data, moving from the blue-shifted velocity component of $\sim$-2\,km\,$\rm s^{-1}$ on the top left to the blue-shifted velocity component $\sim$3.4\,km\,$\rm s^{-1}$ on the bottom right. Contours are at levels 4 to 20\,K\,km\,$\rm s^{-1}$ in steps of 4\,K\,km\,$\rm s^{-1}$.
The velocity channel map of [C\,II] for IC\,59 is shown in Figure \ref{fig:cmap59} with the same general properties as Figure \ref{fig:cmap} but with velocities from -2.5\,km s$\rm^{-1}$ to $\sim$2.9\,km s$\rm^{-1}$, labels on the lower right, and contours at levels 2, 5, and 8\,K\,km\,$\rm s^{-1}$.

\subsection{Position-Velocity Diagrams}

\begin{figure}
\centering
\resizebox{7.5cm}{7cm}{\includegraphics{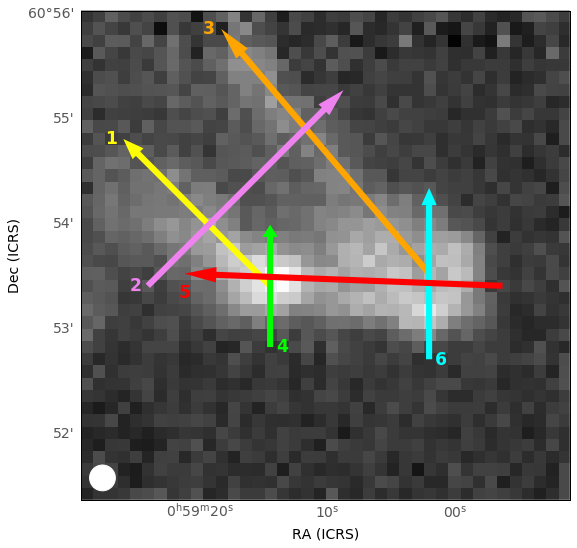}}\\
\resizebox{8.5cm}{6.5cm}{\includegraphics{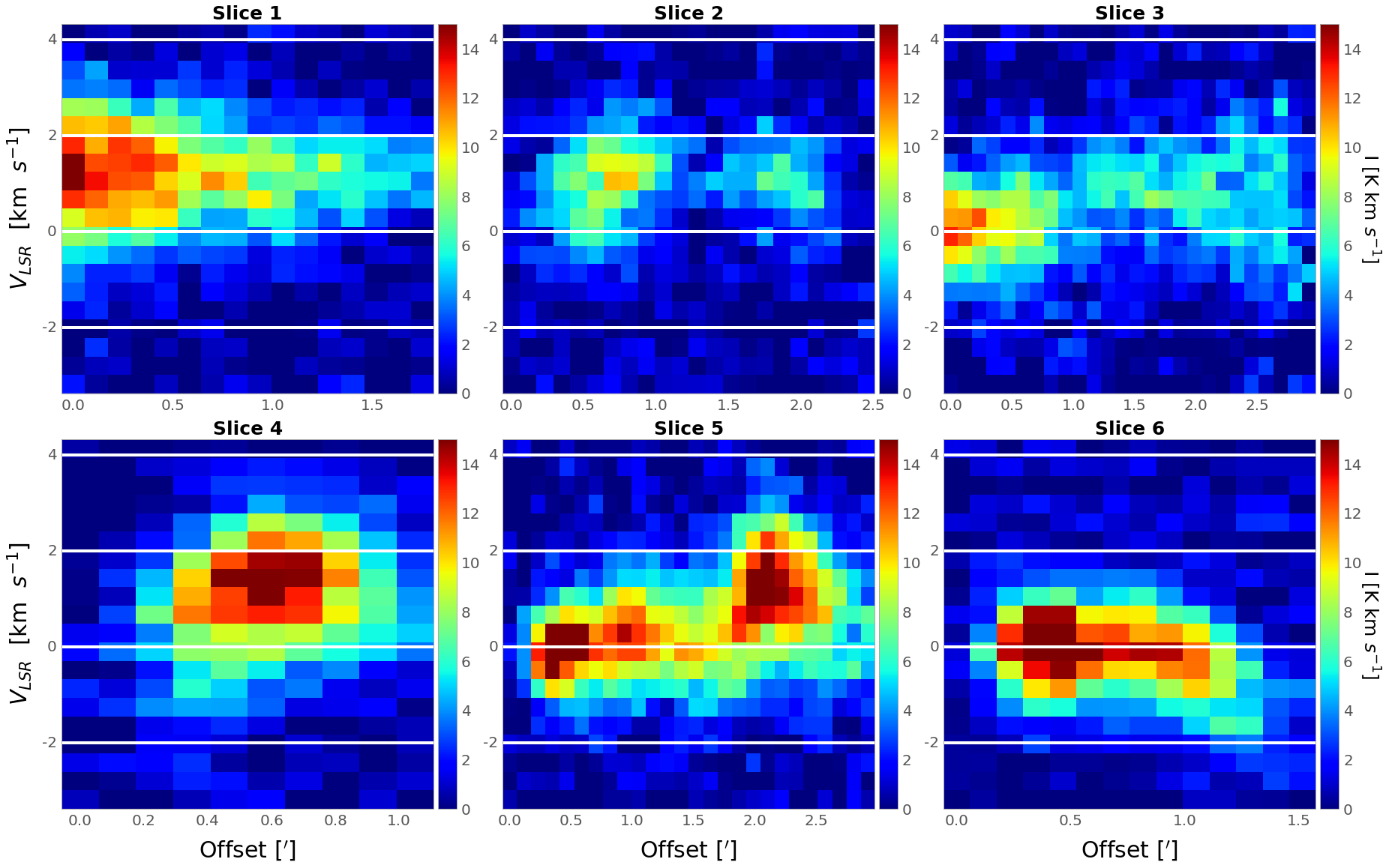}}\\
\resizebox{8.5cm}{3.5cm}{\includegraphics{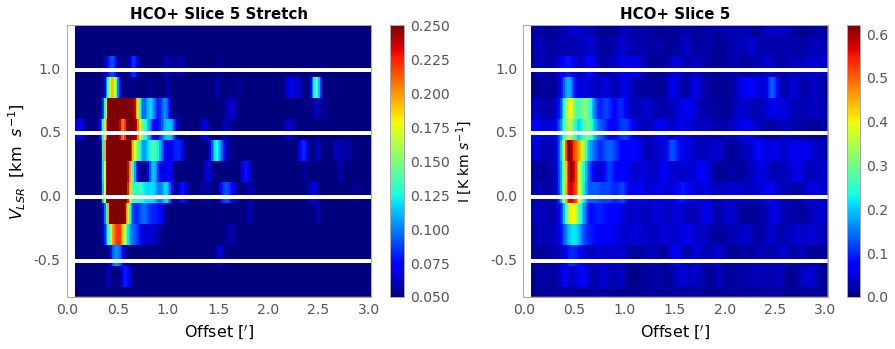}}
\caption{{\bf Upper panel:} gray-scale [C\,II] integrated intensity map of IC\,63 with arrows indicating where and in which direction PV slices were taken.
{\bf Middle panel:} PV diagrams for the six slices, shown in upper plot, are plotted over the same velocity range, from -3 to 4\,km\,s$\rm^{-1}$, and represented by the same color scale.
PV diagrams representing slices along the tails are on the top row while those representing the clumps are on the bottom.
{\bf Lower panel:} PV diagrams for slice 5 in $\rm HCO^{+}$\,(J=1-0)
and formatted like above. The image on the left has had the colors stretched in order to see the the velocity intensity peak around an offset of 2.5$\arcmin$. The image on the right has a more even color map.
\label{fig:pv}}
\end{figure}
\begin{figure}
\centering
\resizebox{8.536cm}{7.58cm}{\includegraphics{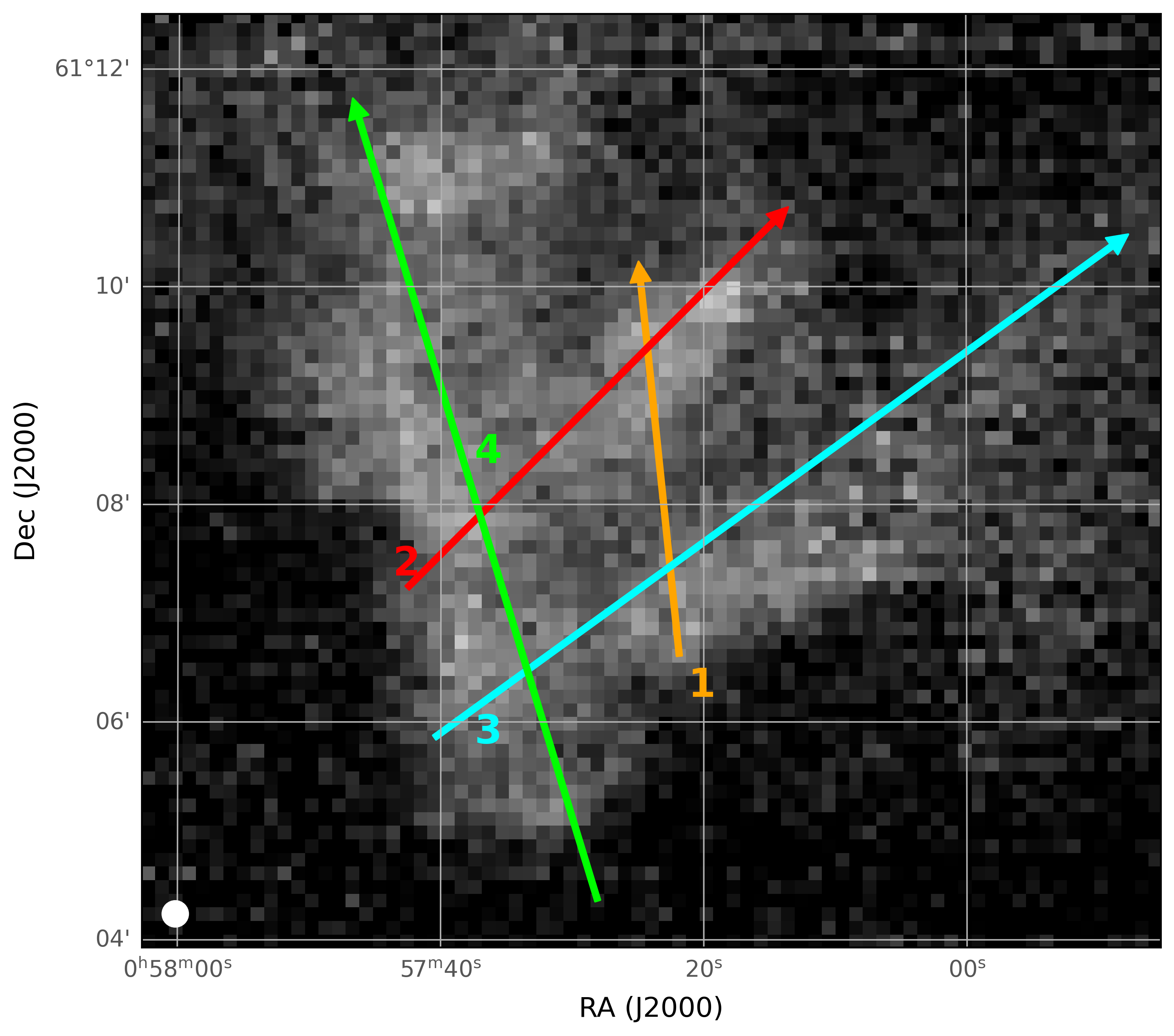}}\\
\resizebox{8.536cm}{7.58cm}{\includegraphics{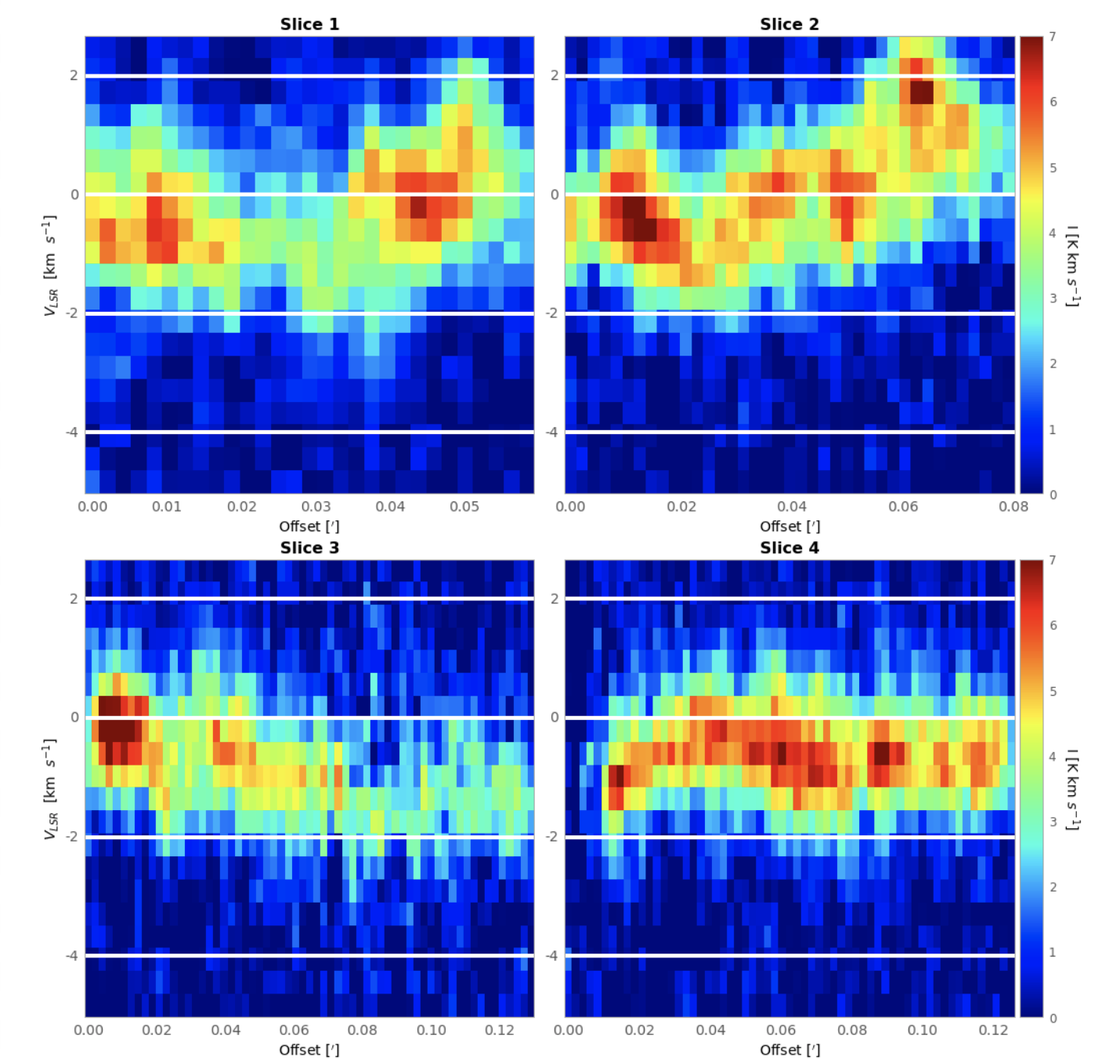}}\\
\caption{Same as Figure \ref{fig:pv} but for IC\,59 and with PV slices plotted from -4 to 3 km\,s$\rm^{-1}$.
\label{fig:pv59}}
\end{figure}

The upper panels of Figures \ref{fig:pv} and \ref{fig:pv59} show gray-scale [C\,II] integrated intensity maps with numbered arrows where slices have been taken through the position-position-velocity upGREAT [C\,II] data cubes of IC\,63 and IC\,59. 
The position-velocity (PV) diagrams from these slices are shown in the lower panels of the figures, with velocities from $\sim$-3 through 4 km\,$\rm s^{-1}$ for IC\,63 and $\sim$-4 through 3 km\,$\rm s^{-1}$ for IC\,59.
$\rm HCO^{+}$\,(J=1-0) data from slice 5 in IC\,63
is also shown on the bottom panel for Figure \ref{fig:pv} from -1 to 1.5\,km\,s$\rm ^{-1}$ in two different color stretches to show the slight emission at an offset of $\sim$2.5$\rm ^{\prime}$.
The offset begins at $\rm 0^{\prime}$ from the base of each arrow from the upper panels and concludes at the tips of the arrow heads.

\section{Results} 
\label{sec:results}

\subsection{Dust and Gas Relations}

\subsubsection{IC\,63}

As can be seen in Figures \ref{fig:cc} and \ref{fig:dust}, the highest intensity regions of IC\,63 are the fishhook feature inside the main clump, the small clump, and the front edge in those maps where the resolution is good enough to see it. 
The other noticeable, but less intense, features are the upper and lower tails that appear to stream from the different clumps. 
The 250\,$\micron$ map from \textit{Herschel} SPIRE, panel (g) in Figure \ref{fig:dust}, has a background source to the west of the PDR that is not seen in any other map shown in Figure \ref{fig:dust}. This object is not being considered in our analysis.

In low-resolution (44$^{\prime\prime}$) observations of IC\,63 using CO (J=1-0) from TRAO, the emission was seen in the main clump, but the small clump was not detected (Soam et al., private communication). The small clump is detected in $\rm HCO^{+}$\,(J=1-0) and [C\,II] emissions .

The dense gasses, like $\rm HCO^{+}$\,(J=1-0), mainly trace the fishhook feature while the ionized gas, traced by H$\alpha$, appears around the areas of the lowest optical depth, such as the western edges of the main and small clumps as well as the upper tail. H$\alpha$ is also present over the majority of the main clump at a slightly lower emission.

The tail features lie slightly off of the radial line from $\gamma$ Cas found to the southwest (see white arrow in Figure \ref{fig:dust} panel (a)).
\citet{2013ApJ...775...84A} argued that, if the angle discrepancy between the projected radial vector from $\gamma$ Cas and the angle of the tails is due to projection effects, the line between the star, $\gamma$ Cas, and the nebula, IC\,63, makes an angle of $\sim$58$^{\circ}$ with respect to the plane of the sky.

\subsubsection{IC\,59}

IC\,59, seen in Figures \ref{fig:cc59} and \ref{fig:dust59}, has multiple regions of high intensities. The brightest region is the upper arm, followed by the main body and the head. The eastern and western regions of the lower arm are the least bright regions. The TRAO observations of $\rm ^{12}CO$\,(1-0) toward IC\,59 only traces the upper arm area, while the H$\alpha$ traces only the bottom ridge of the nebula without overlapping with any of the other tracers, unlike what is seen in IC\,63.  A slight enhancement in the H$_2$ fluorescence can be seen on the star-ward side of the CO detection, reminiscent of the much stronger detection in IC\,63.

\subsection{Cloud Velocities}

\subsubsection{IC\,63}
One of the most noticeable features of Figure \ref{fig:cmap} is the intensity shift from west to east as the velocities move from negative to positive.
The most blue shifted velocity region of the nebula is the western edge of the main clump, including the fishhook and front edge, which also happens to be where H$\alpha$ and $\rm H_{2}$\,(1-0)\,S(1) are dominant (see Figure \ref{fig:cc}). 
This region disappears in the more positive velocities, where the rest of the cloud is visible.

As can also be seen with the PV diagrams in Figure \ref{fig:pv}, the small clump, slice 4, seems to span the largest velocity ranges from $\sim$-1\,km\,s$\rm^{-1}$ through 3\,km\,s$\rm^{-1}$ while the main clump/fishhook, slice 6, seems to only range from $\sim$-2\,km\,s$\rm^{-1}$ through 1\,km\,s$\rm^{-1}$.
The wide range in velocities covered by the small clump leads to a larger integrated intensity, as seen in Figure \ref{fig:dust} panel (a), even though the peak intensities never surpass those of the fishhook.
This could mean that the small clump is less dense and cooler than the fishhook region.

Slice 5 in Figure \ref{fig:pv} shows that the main clump region, including the fishhook and front edge, is much more blue shifted in velocity than the small clump, with average velocities of $\sim$0 and 1.5 km\,s$\rm^{-1}$, respectively. The location of the $\rm HCO^{+}$\,(J=1-0) emission coincides with the emission of [C\,II].

Slice 2 shows that the tails are somewhat blue shifted in velocity along the outer edges of the nebula, where some H$\alpha$ is seen in Figure \ref{fig:cc}, while being redshifted towards the area in between the two tails.

\subsubsection{IC\,59} \label{section:59kin}

The velocity channel map of [C\,II] in IC\,59, shown in Figure \ref{fig:cmap59}, seems to follow similar patterns to IC\,63 in Figure \ref{fig:cmap}, but in opposite velocity directions. 
The majority of the cloud is blue shifted in velocity and clearly visible until -0.5\,km\,s$\rm^{-1}$ where it begins to disappear. 
The main body, eastern lower arm, and the upper arm are the only regions that persist into the positive velocities. The highest intensity region, the upper arm, goes $\sim$1\,km\,s$\rm^{-1}$ further into the positive velocities than any other region.

The upper arm of IC\,59, where CO is detected and shown in blue in the upper panel of Figure \ref{fig:cc59}, is comparable to the main clump/fishhook region of IC\,63 because these are the densest regions of each cloud.  This is also where a faint H$_2$ fluorescent ridge can be seen.
Comparing these two high density regions in each cloud we find a contrast in their movements compared to the rest of the clouds. The high density region in IC\,63 contains, and is surrounded by, the most negative [C\,II] velocities seen in the cloud while we see the opposite in IC\,59 with the most positive velocities seen in and around the high density region of the upper arm.

Slice 1 of Figure \ref{fig:pv59} shows that the outside of the two arms are redshifted in velocity, while the space between them is blue shifted. 
In comparison, slice 2 of Figure \ref{fig:pv} shows that IC\,63 is the opposite, where the outside of the tails are blue shifted and the area in between is redshifted. 
This suggests that the ionized gas towards the outside of both the arms and tails are traveling in opposite directions from our perspective.

\subsection{Spectral line features in different regions of the nebulae}

\begin{figure}
\centering
\resizebox{5cm}{4.5cm}{\includegraphics{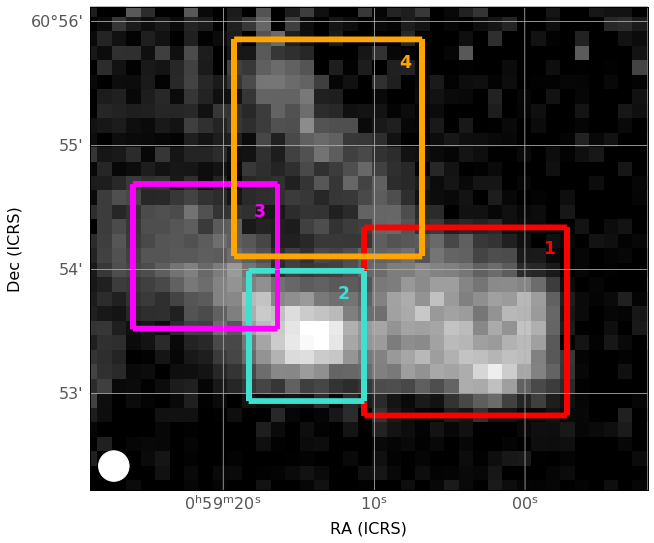}}\\
\resizebox{8cm}{7cm}{\includegraphics{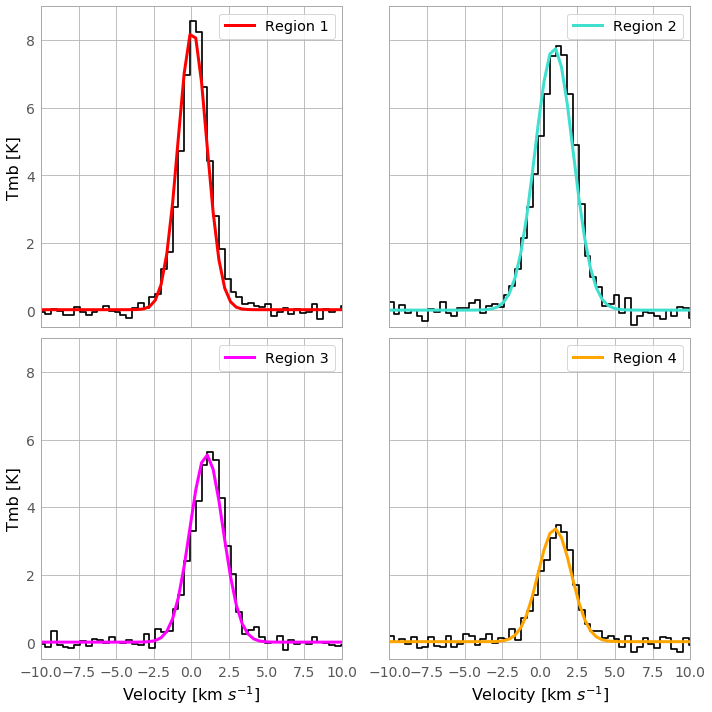}}\\
\resizebox{8cm}{7cm}{\includegraphics{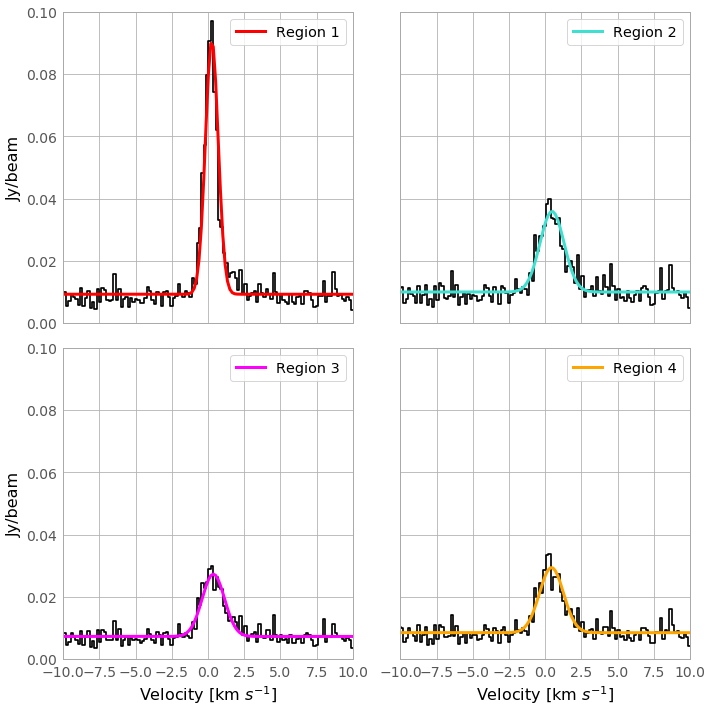}}
\caption{{\bf Upper panel:} boxed boundaries for the four main regions of IC\,63 overlaid on the [C\,II] integrated intensity map, shown in gray-scale. {\bf Middle panel:} averaged line extraction spectra of [C\,II], black, from each of the four regions from above with a Gaussian fitted in the same color as the region box it represents. {\bf Lower panel:} averaged line extraction spectra from $\rm HCO^{+}$\,(J=1-0) over the same regions and colors as [C II].
Fit parameters for each Gaussian are shown in Table \ref{table:1}}.\label{fig:lines}
\end{figure}

\begin{figure}
\centering
\resizebox{8.5cm}{8cm}{\includegraphics{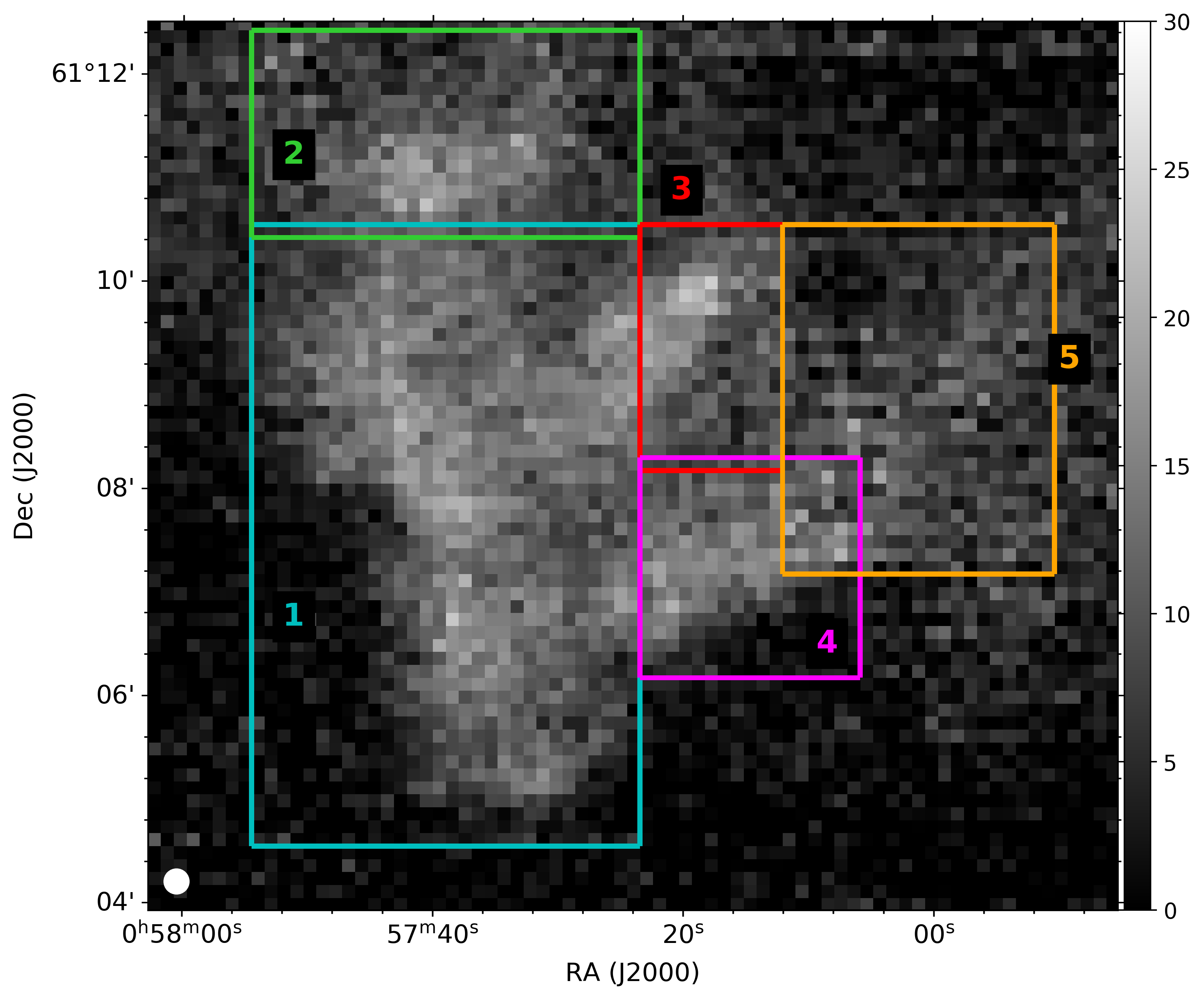}}\\
\resizebox{8.5cm}{8cm}{\includegraphics{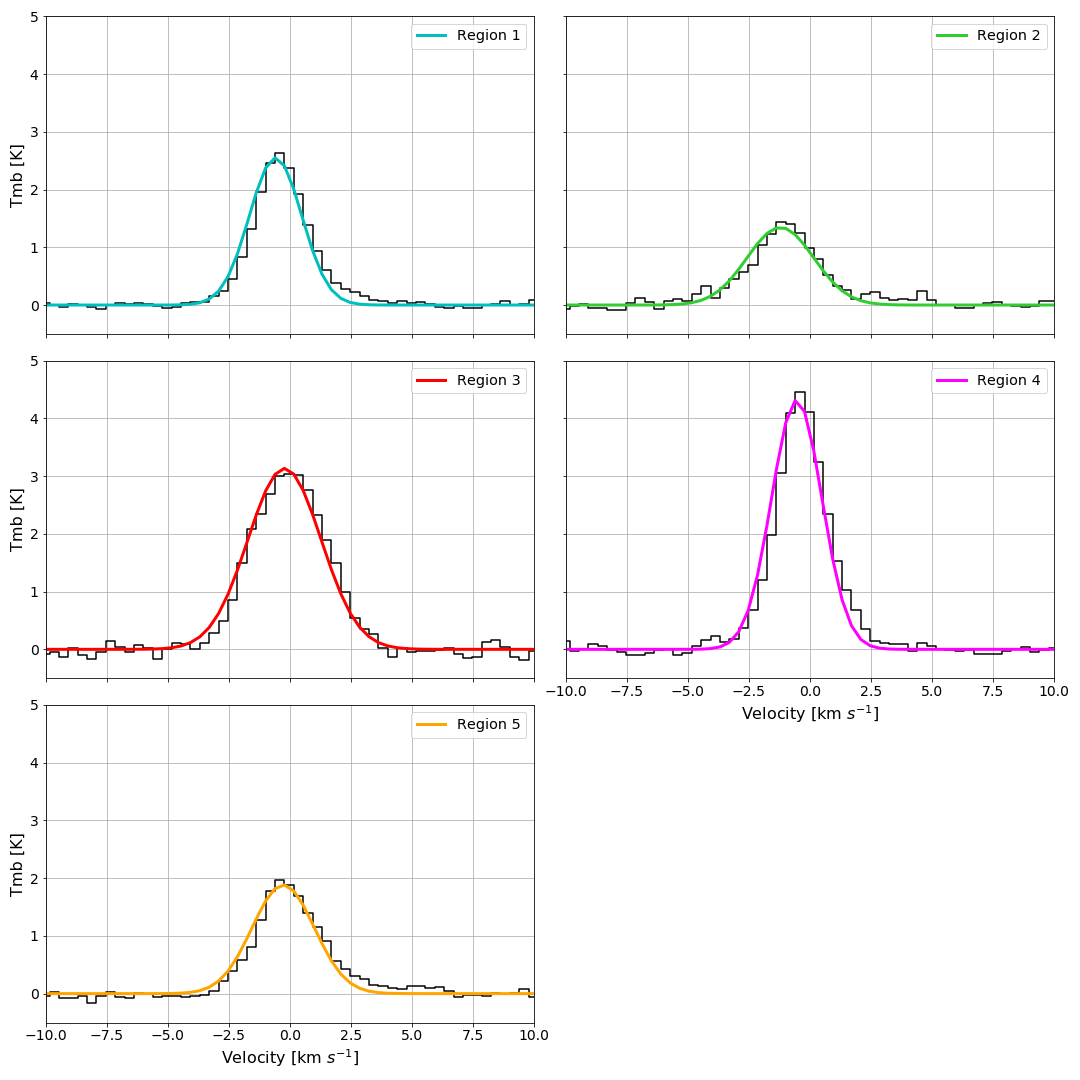}}\\
\caption{Same as Figure \ref{fig:lines} but for  five regions in IC\,59 and with extracted spectra filled in gray. Fit parameters for each Gaussian are shown in Table \ref{table:2}.}\label{fig:lines59}
\end{figure}

The lower panels of Figures \ref{fig:lines} and \ref{fig:lines59} show spectra that have been extracted from the SOFIA/upGREAT [C\,II] position-position-velocity cube regions for IC\,63 and IC\,59, respectively, along with the spectra for $\rm HCO^{+}$\,(J=1-0) from these same regions in IC\,63. The regions are over plotted and numbered on the gray-scale [C\,II] integrated intensity maps in the upper panels of each figure.
Each of the spectra in the lower panels have been fitted with a single Gaussian function to derive their quantitative properties, listed in Tables \ref{table:1} and \ref{table:2} for IC\,63 and IC\,59, respectively.

\begin{table}
\centering
\caption{Table of the parameters for the Gaussian fitted to the [C\,II] and $\rm HCO^{+}$\,(J=1-0) velocity spectra of IC\,63 for each region in Figure \ref{fig:lines} 
}\label{table:1}
\scriptsize
\begin{tabular}{llllc}\hline
Region & $\rm V_{peak}\pm \sigma_{V_{peak}}$ & $ FWHM \pm \sigma_{FWHM}$ & $\rm T_{mb}\pm \sigma_{T_{mb}}$ \\
C\,II & (km s$\rm^{-1}$)  & (km s$\rm^{-1}$)    & (K)         \\\hline
1 & 0.06 $\pm$ 0.01 & 2.2 $\pm$ 0.03 & 8.28 $\pm$ 0.10 \\ 
2 & 0.95 $\pm$ 0.02 & 3.0 $\pm$ 0.05 & 7.77 $\pm$ 0.11 \\
3 & 1.0 $\pm$ 0.02 & 2.6 $\pm$ 0.06 & 5.55 $\pm$ 0.10 \\
4 & 1.0 $\pm$ 0.03 & 2.6 $\pm$ 0.08 & 3.36 $\pm$ 0.09 \\
\hline
Region & $\rm V_{peak}\pm \sigma_{V_{peak}}$ & $ FWHM \pm \sigma_{FWHM}$ & $\rm I_{peak}\pm \sigma_{I_{peak}}$ \\
$\rm HCO^{+}$      &(km s$\rm^{-1}$)       &(km s$\rm^{-1}$)              & (Jy/beam)         \\\hline
1 & 0.27 $\pm$ 0.01 & 1.04 $\pm$ 0.03 & 0.082 $\pm$ 0.002 \\ 
2 & 0.47 $\pm$ 0.05 & 1.96 $\pm$ 0.11 & 0.026 $\pm$ 0.001 \\
3 & 0.37 $\pm$ 0.04 & 1.81 $\pm$ 0.10 & 0.020 $\pm$ 0.001 \\
4 & 0.43 $\pm$ 0.05 & 1.96 $\pm$ 0.12 & 0.021 $\pm$ 0.001 \\
\hline
\hline
\end{tabular}
\end{table}

\begin{table}
\centering
\caption{Table of the parameters for the Gaussians fitted to the [C\,II] velocity spectra of IC\,59 for each region in Figure \ref{fig:lines59}.}\label{table:2}
\scriptsize
\begin{tabular}{llllc}\hline
Region & $\rm V_{peak}\pm \sigma_{V_{peak}}$ & $ FWHM \pm \sigma_{FWHM}$ & $\rm T_{mb}$ \\
C\,II      &(km s$\rm^{-1}$)       &(km s$\rm^{-1}$)              & (K)         \\\hline
1 & -0.58 $\pm$ 0.01 & 2.5 $\pm$ 0.03 & 2.6 $\pm$ 0.03 \\ 
2 & -1.2 $\pm$ 0.05 & 3.2 $\pm$ 0.11 & 1.3 $\pm$ 0.04 \\
3 & -0.21 $\pm$ 0.02 & 3.5 $\pm$ 0.05 & 3.1 $\pm$ 0.04 \\
4 & -0.53 $\pm$ 0.01 & 2.4 $\pm$ 0.03 & 4.3 $\pm$ 0.05 \\
5 & -0.26 $\pm$ 0.03 & 3.0 $\pm$ 0.06 & 1.9 $\pm$ 0.04 \\
\hline
\end{tabular}
\end{table}

The differences in peak $\rm T_{mb}$ between the two clouds is very obvious in these tables, with the most intense regions in IC\,59 having about the same peak intensity as the two weakest regions in IC\,63. 
The central velocities also vary, with IC\,63 going from $\sim$0\,to \,1\,km\,s$\rm^{-1}$ and IC\,59 ranging from -0.5\,to \,-1.2\,km\,s$\rm^{-1}$.

Most of the spectra for IC\,63 in Figure \ref{fig:lines} show a \textbf{slight, but systematic asymmetry, with a more} gradual rise on the blue side and a steeper drop off on the red for [C\,II] and a fairly uniform shape for $\rm HCO^{+}$\,(J=1-0).

\subsection{Velocity Gradients and Magnetic Fields}

\begin{figure}
\resizebox{8.5cm}{7.0cm}{\includegraphics{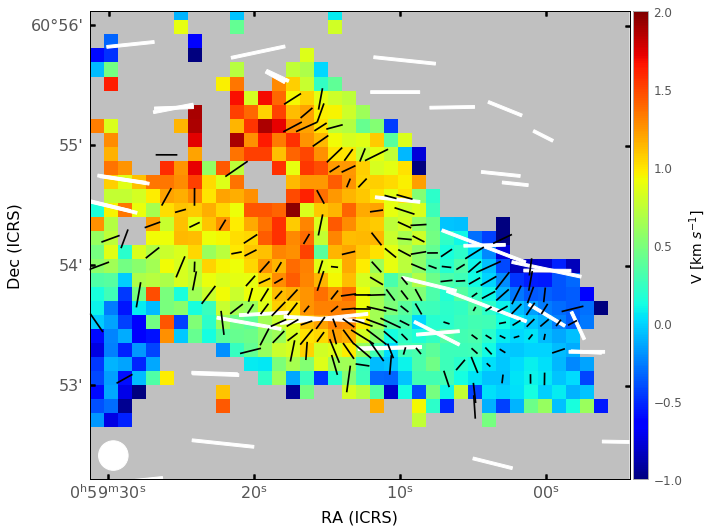}}
\resizebox{8.5cm}{7.0cm}{\includegraphics{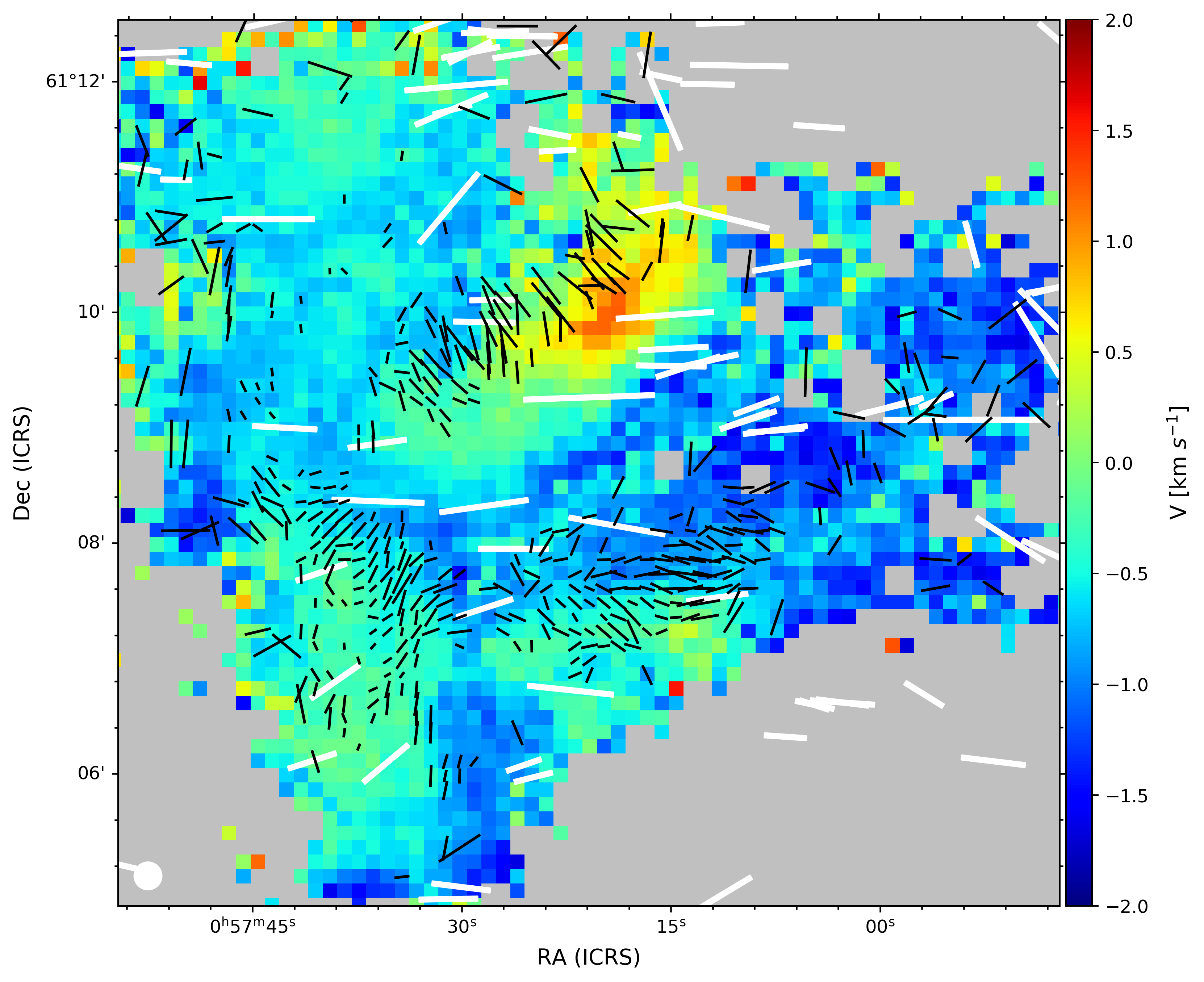}}
\caption{Central velocity maps of IC\,63, top, and IC\,59, bottom, with their respective velocity gradient vectors plotted in black. Magnetic field polarization vectors are plotted in white on both maps and represent data from \citet{2013ApJ...775...84A,2017MNRAS.465..559S} for IC\,63 and \citet{2017MNRAS.465..559S} for IC\,59. The length of the lines are representative of the magnitude of the velocity changes for the velocity gradients and the polarization percentages for the magnetic field polarizations in both maps. }\label{fig:vgradpol}
\end{figure}

In order to understand any relation between gas velocity and magnetic fields, we calculated velocity gradient of [C\,II] emission and compared those with magnetic field orientations taken from \citet{2017MNRAS.465..559S}.
These gradients are not absolute velocities,  but indicate local accelerations.
Gradients were calculated by taking the difference between the central velocities - estimated through fitting each pixel's velocity spectrum with a Gaussian - of the pixels a specified distance from a central pixel.
Only velocity gradients with less than a 20$^{\circ}$ error in angle are used. 
Different step distances, 2 vs. 4 pixels, were used for IC\,63 and IC\,59 respectively, to best show the dynamics for the clouds. 
A larger step size was needed for the gradients in IC\,59 due to the cloud being much larger than IC\,63 and the pixel-to-pixel velocity change being smaller.

Figure \ref{fig:vgradpol} shows the local velocity gradient vectors of IC\,63 and IC\,59 overlaid in black on their respective central velocity maps. Previously published magnetic field orientations from \citet{2013ApJ...775...84A, 2017MNRAS.465..559S} for both nebulae, obtained from observations of partially extinguished starlight from stars located behind the nebula, are plotted with white line segments. It is hard to draw any firm conclusions on the relative orientations of the velocity gradients and magnetic field lines. However, there are some hints of the magnetic fields being perpendicular to the velocity gradients in both the nebulae. High-resolution polarization observations in longer wavelengths may shed more light on this investigation.

\section{Discussion} 
\label{sec:discussion}

\subsection{FUV Radiation Field at IC\,63}   \label{ic63uv}

The FUV radiation field ($\rm G_{0}$) incident on IC\,63 from $\gamma$ Cas has been estimated by various previous studies.
\citet{1995A&A...302..223J}, using the projected distance between $\gamma$ Cas and IC\,63 of 1.3\,pc and a distance to $\gamma$ Cas of 230\,pc, found that $\rm G_{0}$ at IC\,63 was 650 units of the Draine field \citep{1978ApJS...36..595D}.
\citet{2018A&A...619A.170A} used dust emissions in IC\,63 instead of spectral emissions from $\gamma$ Cas and found a $\rm G_{0}$ of 150, indicating that the nebula was further away than what  \citet{1995A&A...302..223J} assumed as the projected distance. 
\citet{2013ApJ...775...84A} proposed that, based on the angle of the streaming tails in IC\,63, the cloud lies behind $\gamma$ Cas at an angle of 58$^{\circ}$, making the true distance between $\gamma$ Cas and IC\,63 $\sim$2\,pc.

Since $\gamma$ Cas is a variable star, comparing the FUV spectra from \citet{1979ApJS...39..195C}, used by \citet{1995A&A...302..223J}, to a newer spectrum from \citet{2005ApJ...628..750F}, gives a better idea of what the star may have been like during the measurements of IC\,63 shown by \citet{2018A&A...619A.170A} as well as systematic errors in the processes used.
It is found that the flux from the \citet{2005ApJ...628..750F} spectra is $\sim$.65 $\times$ the flux found in the work by \citet{1979ApJS...39..195C}.
Using both of these spectra, the newer parallax to $\gamma$ Cas, and a distance from $\gamma$ Cas to IC\,63 of 2\,pc, we find that $\rm G_{0}$ at IC\,63 is 115 and 175 from \citet{2005ApJ...628..750F} and \citet{1979ApJS...39..195C}, respectively.

\subsection{FUV Radiation Field at IC\,59}

\begin{figure}
\centering
\resizebox{8.57cm}{3.71cm}{\includegraphics{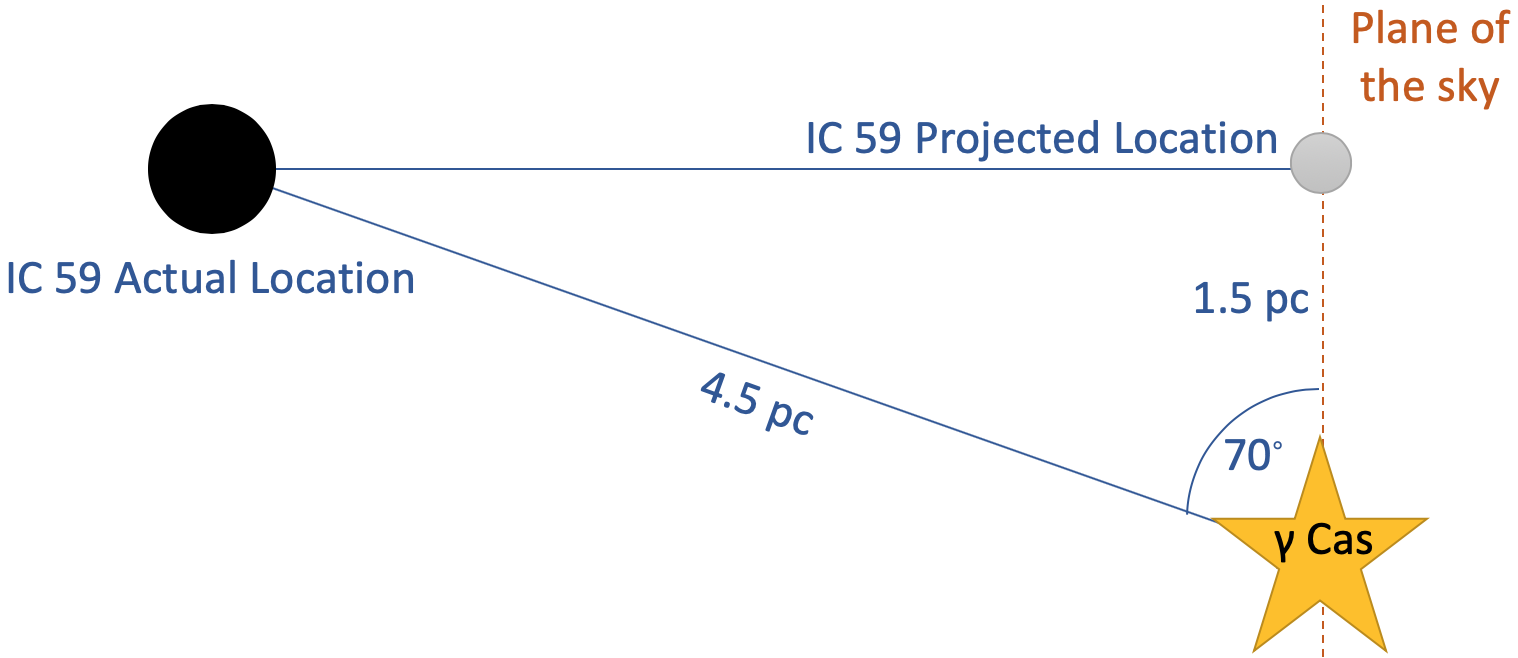}}
\caption{Side view model of the IC\,59 - $\gamma$ Cas system with geometrical findings based on $\rm G_{0}$ calculations.}\label{fig:G0math}
\end{figure}

We are unable to estimate a $\rm G_{0}$ for IC\,59 in the same way as we did in Section \ref{ic63uv} for IC\,63, because IC\,59 has a mostly vertical orientation in the plane of the sky which would mitigate any projection effects from an inclination angle. 
Using the projected distance from $\gamma$ Cas to IC\,59 of 1.5\,pc provides a $\rm G_{0}$ that is far higher than the 25 found by \citet{2018A&A...619A.170A}. There may be a possibility of leakage of factors FUV radiation field due to clumpiness in the clouds. In order to find the inclination angle for IC\,59, we will assume that the $\rm G_{0}$ from \citet{2018A&A...619A.170A} and our methods used in Section \ref{ic63uv} are correct because our calculated $\rm G_{0}$ for IC\,63 agrees, within statistical errors, with the 150 that \citet{2018A&A...619A.170A} found.

Utilizing these assumptions, we calculate that a distance of 4.5\,pc from $\gamma$ Cas to IC\,59 will produce a $\rm G_{0}$ of 25. 
This distance corresponds to an inclination angle of 70$^{\circ}$ with respect to the plane of the sky.
Figure \ref{fig:G0math} shows the proposed geometry for the IC\,59\,-\,$\gamma$ Cas system as viewed from the side.

\subsection{Kinetic vs. Magnetic Energy}

In IC\,63, Figure \ref{fig:vgradpol} upper panel, [C\,II] velocity gradients show a fairly connected system that moves from velocities of almost 0\,km\,s$^{-1}$ at the main clump into the velocities of 1\,km\,s$^{-1}$ or more in the tails. No obvious structural correlation could be found when comparing the local velocity gradients and the magnetic field orientations at IC\,63.

In IC\,59, the data do not show a connected system (see lower panel of Figure \ref{fig:vgradpol}). There is slightly more anti-correlation observed between the local velocity gradients and the magnetic field orientations in IC\,59 than what was observed in IC\,63. As described in \citet{2017MNRAS.465..559S}, this may be related to the concave shape created by UV radiation on the star-facing side. The field lines follow this concave shape of the cloud rather than going along the radiation direction as seen in IC\,63

\citet{2017MNRAS.465..559S} found the dynamical pressure in IC\,63 to be less than the magnetic pressure, and both were less than the external pressure for IC\,63. Using their magnetic field strength average over the main body of the PDR (region II in their Figure 1) and an average velocity change in our velocity gradients, we estimate magnetic and kinetic energy values of $\rm E_{mag}$ = $\rm 3.4 x 10^{32}~J$ and $\rm E_{kin}$ = $\rm 3.8 x 10^{42}~J$, respectively, which suggests that the kinetic energy dominates the magnetic energy in IC\,63.

\subsection{Geometry of the System}

\begin{figure}
\resizebox{8cm}{7.5cm}{\includegraphics{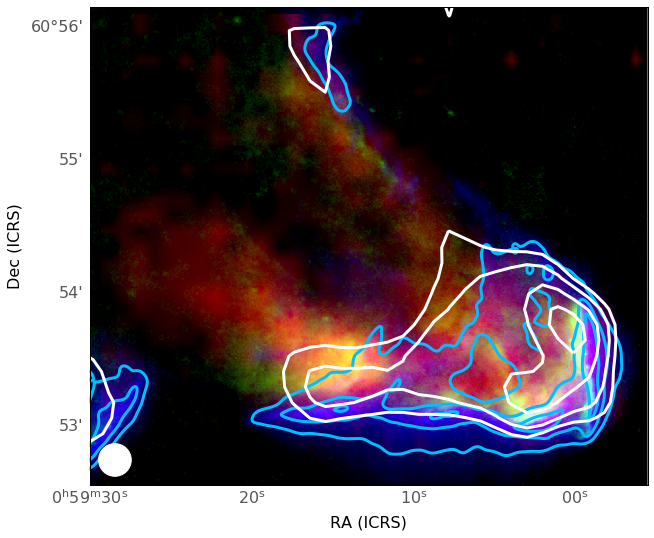}}
\caption{RGB from upper panel of Figure \ref{fig:cc} with smoothed contours of [C\,II] integrated intensity below -0.5\,km\,s$\rm^{-1}$ in white with levels [5,7,10,13]\,K\,km\,$\rm s^{-1}$ and smoothed H$\alpha$ contours in blue with levels [150,200,300] pixel counts.
The beam size for the SOFIA [C\,II] observations, white contour, is plotted in the bottom left corner. 
}\label{fig:Ha_RGB_cont}
\end{figure}

\begin{figure}
\resizebox{8.3cm}{6.71cm}{\includegraphics{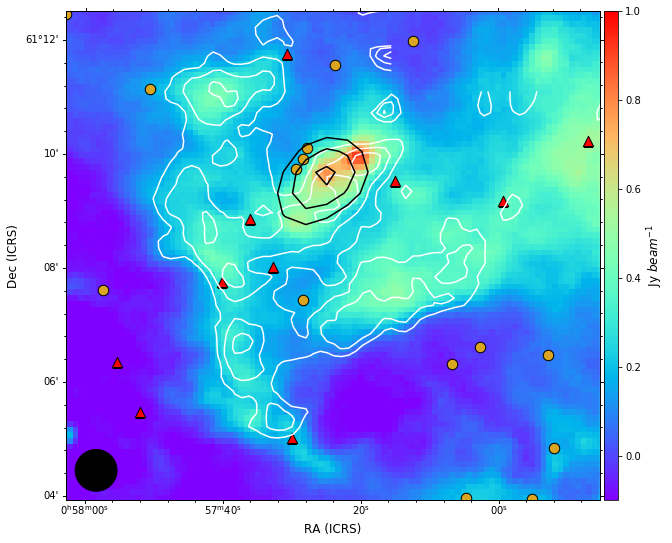}}
\caption{250 $\micron$ dust map of IC\,59 with C [II] integrated intensity contours at levels [10, 13, 16, 20]\,K\,km\,$\rm s^{-1}$ in white for visual reference and $\rm ^{12}CO$\,(1-0) contours at levels [1500, 2500, 3300]\,K\,km\,s$\rm^{-1}$ in black. Markers are stars from the VILNIUS Survey within 1250\,pc with an $\rm A_{v}$ less than 1 for the yellow circles and greater than 1 for the red triangles. Beam size of the $\rm ^{12}CO$\,(1-0) contours is shown in black in bottom left corner.
}\label{fig:Ha_RGB_cont59}
\end{figure}

As we have found that each nebula lies off the plane of the sky due to actual distances from the stars to the nebula being larger than the projected distances, we attempted to find whether the nebulae were in front of or behind $\gamma$ Cas.

For IC\,63, the RGB images in Figure \ref{fig:cc}, the shifting features seen in the channel map in Figure \ref{fig:cmap}, the PV diagrams in Figure \ref{fig:pv}, and the velocity gradients in Figure \ref{fig:vgradpol}, indicate that the blue shifted [C\,II] gas, where $\rm v < \,0\,km\,s^{-1}$, is correlated with the H$\alpha$ emission that is likely due to photo-evaporation. This correlation is shown in Figure \ref{fig:Ha_RGB_cont}, where the same RGB image as the middle panel of Figure \ref{fig:cc}, with red as [C\,II] integrated intensity, green as $\rm H_{2}$\,(1-0)\,S(1), and blue as H$\alpha$, has been plotted with H$\alpha$ contours in blue and contours of the [C\,II] integrated intensity map of velocities below -0.1\,km\,s$\rm^{-1}$ are plotted in white. These two contours correlate well, with the ionized H$\alpha$ sitting just outside the cloud of the blue shifted velocities of [C\,II].
Also, as shown in Table \ref{table:1}, while the [C\,II] central velocities range from 0\,-\,1\,km\,s$\rm ^{-1}$ the $\rm HCO^{+}$\,(J=1-0) velocities only range from $\sim$ 0.27\,-\,0.47\,km\,s$\rm ^{-1}$.
This suggests that these two emissions represent the same region and that the gasses in this region are both being actively ionized and blue shifted while the rest of the cloud is not. This suggests that IC\,63 lies on the far side of $\gamma$ Cas. Our finding is consistent with findings in \citet{2013ApJ...775...84A}.

Similarly, the RGB images of IC\,59 in Figure \ref{fig:cc59}, the channel map in Figure \ref{fig:cmap59}, the PV diagrams in Figure \ref{fig:pv59}, and the velocity gradients in Figure \ref{fig:vgradpol}, show that the H$\alpha$ emission does not overlap with the dust or gas emissions.
Because the H$\alpha$ emission in IC\,59 does not overlap with the rest of the dust or gas emissions, we consider a slightly differently approach than for IC\,63. The locations that are closest to the H$\alpha$ emission are redshifted in IC\,59, with v $>$ 0\,km\,s$\rm ^{-1}$, indicating that the emission could be coming from the far side of the cloud. The H$\alpha$ emission coincident with the [C\,II] emission would then likely be extincted by the cloud.
Figure \ref{fig:Ha_RGB_cont59} shows the 250 $\micron$ dust map for IC\,59 with [C\,II] integrated intensity contours in white and $\rm ^{12}CO$\,(1-0) contours in black. The filled circles and triangles indicate the locations of stars less than 1.25\,kpc distance with measures of visual extinctions $\rm A_V$ \citep{soam_poleff2021}. Yellow circles show stars with an $\rm A_V$ $<$ 1\,mag, while red triangles indicate stars with an extinction $>$ 1\,mag.

The differences in the relative velocities of the high density regions of the two clouds, along with the opposing relative velocity differences of the diffuse gas emissions in between the arms of each cloud (Section \ref{section:59kin}) support that the IC\,59 and IC\,63 are on opposite sides of $\gamma$ Cas. We therefore predict that IC\,59 is on the near side of $\gamma$ Cas.

\begin{figure}
\centering
\resizebox{6cm}{5.5cm}{\includegraphics{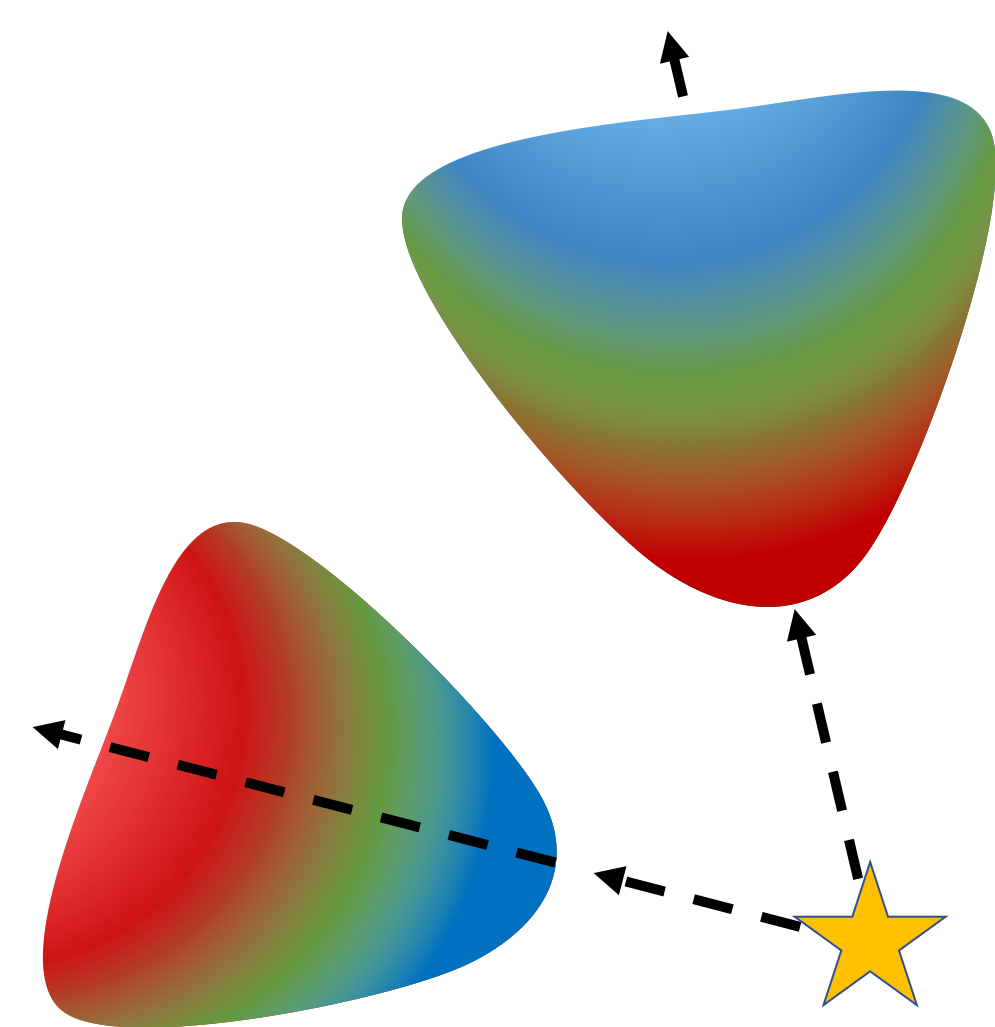}}
\caption{Cartoon model of the $\gamma$ Cas, IC\,59, and IC\,63 system. Dashed black lines represent the plane of the sky at the distance of $\gamma$ Cas.}\label{fig:model}
\end{figure}

Our proposed geometric model of the Sh 2-185 system is shown in a simplified form in Figure \ref{fig:model}, where the nebulae are modeled by red, green, and blue colors representing positive, zero, and negative velocities respectively. The dashed lines from the star indicate the plane of the sky containing $\gamma$ Cas.

\section{Summary} \label{sec:summary}
We studied the structure and kinematics of PDRs IC\,59 and IC\,63 associated with H\,II region Sh 2-185 using high-resolution SOFIA/upGREAT [C\,II] observations combined with archival data on dust emission toward these nebulae. The case of IC\,59 and IC\,63 presents an interesting example in which the feedback from a B0 star is dominated
by photevaporation.
\begin{enumerate}
    \item The [C\,II] emission of IC\,63 has two clumps with streaming tails showing a velocity shift towards positive velocities, moving away from $\gamma$ Cas. IC\,59 has a less defined structure, consisting of multiple low density regions and one large core that is redshifted in comparison to the rest of the cloud, which otherwise shift from positive to negative velocities when going away from $\gamma$ Cas.
    \item A $\rm G_{0}$ of 115-175 was found for IC\,63 using a previously determined inclination angle of 58$^{\circ}$ \citep{2013ApJ...775...84A}. Utilizing a $\rm G_{0}$ for IC\,59 of 25 \citep{2018A&A...619A.170A}, a distance of 4.5\,pc and inclination angle of 70$^{\circ}$ could be estimated for the cloud.
    \item Using [C\,II] data, we constructed velocity gradients and compared these to plane-of-the-sky magnetic field orientations. We find an anti-correlation between the velocity gradient and magnetic fields in IC\,59, but no clear structure in IC\,63. The upper limits of our estimations showed that kinetic energy is larger than the magnetic energy in IC\,63. This suggests that kinetic pressure in this nebula is dominant.
    \item  From our analysis of the kinematics and structure of these clouds, we find that IC\,63 lies behind $\gamma$ Cas and IC\,59 lies in front.
\end{enumerate}

\vspace{2.3pt}
\vspace{2.3pt}
\section{Acknowledgements} \label{sec:acknowledgements}

Based in part on observations made with the NASA and DLR Stratospheric Observatory for Infrared Astronomy (SOFIA). SOFIA is jointly operated by the Universities Space Research Association, Inc. (USRA), under NASA contract NNA17BF53C, and the Deutsches SOFIA Institut (DSI) under DLR contract 50 OK 0901 to the University of Stuttgart. The GREAT team was an invaluable resource in obtaining and utilizing the data for this project.

Financial support for this work was provided by NASA through award 05\_0052 issued by USRA. A.S. and B-G.A. are supported by National Science Foundation Grant-1715876. This material is also based upon work supported by the National Science Foundation under Grant No. 1715060. 

This paper makes use of data obtained as part of the INT Photometric H$\alpha$ Survey of the Northern Galactic Plane (IPHAS, \url{www.iphas.org}) carried out at the Isaac Newton Telescope (INT). The INT is operated on the island of La Palma by the Isaac Newton Group in the Spanish Observatorio del Roque de los Muchachos of the Instituto de Astrofisica de Canarias. All IPHAS data are processed by the Cambridge Astronomical Survey Unit, at the Institute of Astronomy in Cambridge. The bandmerged DR2 catalogue was assembled at the Centre for Astrophysics Research, University of Hertfordshire, supported by STFC grant ST/J001333/1.

%
\textit{Facility:} upGREAT/SOFIA
\textit{Softwares:} Astropy \citep{2013A&A...558A..33A}, StarNet++, CLASS



\bibliography{ic6359}{}
\bibliographystyle{aasjournal}

\appendix

\section{Thermal Pressure}

Thermal pressures of these PDRs can be calculated with:
\begin{equation}\label{equation_8}
    {\rm \frac{P_{[C\, II]}}{ k_{B}} = n_{[C\, II]}T_{ex}}
\end{equation}

where $\rm n_{[C\, II]}$ is estimated from column density assuming the prominent $\rm [C\, II]$ region is a cylinder projected as a rectangle, and we assume that the [C\,II] excitation is thermalized. In IC\,63 this cylinder is estimated to have a length of 0.15\,pc and radius of 0.05\,pc. This gives a one sigma limit of $\rm P_{[C\, II]}$/$\rm k_{B}$ to be 210.6\,K $\rm cm^{-3}$. 
If we use $\rm n_{H_{2}}$ = 5 $\pm 2$ $\times$ $\rm 10^{4}$ $\rm cm^{-3}$ \citep{1994A&A...282..605J} and $\rm T_{kin}$ = 106 $\pm 11$ K \citep{2009MNRAS.400..622T}, then we get a  $P_{H_{2}}$/$k_{B}$ of $\rm 53\times 10^{6}$\,K\,$\rm cm^{-3}$.
The $\rm [C\, II]$/$\rm H_{2}$\,(1-0)\,S(1) ratio, using our calculated $\rm n_{[C\, II]}$ and the $\rm n_{H_{2}}$ from \citet{1994A&A...282..605J}, is found to be 1.1 $\times$ $\rm 10^{-5}$. This is about a factor of 10 off of the abundance ratios calculated for the horse head nebula in \citet{2018AJ....155...80B}. This could be partially explained by a difference in volume estimations between our work and \citet{1994A&A...282..605J}.
Using equation \ref{equation_8} for [C\,II] in an area of IC\,59 with a length of 0.315\,pc and radius of 0.101\,pc, the one sigma limit of $\rm P_{[C\, II]}$/$\rm k_{B}$ is found to be 29.19\,K\,$\rm cm^{-3}$.

\end{document}